%% file: BSF_ScalarMed_arXiv_v3.tex
\documentclass[a4paper,11pt]{article}
\pdfoutput=1
\usepackage{jheppub}

\usepackage{amssymb,amsmath,amsthm,amsfonts}
\usepackage{mathabx,mathtools,siunitx,slashed}
\usepackage{graphicx,enumerate,comment}
\usepackage{relsize,mdframed,multirow}
\numberwithin{equation}{section} 
\input{TIKZ_prelim}

\providecommand{\abs}[1]{\lvert#1\rvert} 
\usepackage{cleveref} 

\def\BSFone{{\rm \mathsmaller{BSF1}}}
\def\BSFtwo{{\rm \mathsmaller{BSF2}}}
\def\SC{{\rm \mathsmaller{SC}}}
\def\TC{{\rm \mathsmaller{TC}}}
\def\vrel{v_{\rm rel}}

\preprint{Nikhef-2018-039}

\title{Dark matter bound states via emission of scalar mediators}	

\author{Ruben Oncala and Kalliopi Petraki}

\affiliation{
\href{https://www.nikhef.nl/en/}{\color{black} Nikhef}, 
Science Park 105, 1098 XG Amsterdam, The Netherlands
}

\bigskip

\affiliation{
\href{http://www.lpthe.jussieu.fr/spip/index.php}{\color{black}
Laboratoire de Physique Th\'eorique et Hautes Energies (LPTHE)}, \\
UMR 7589 CNRS \& Sorbonne Universit\'e, 
4~Place Jussieu, F-75252, Paris, France
}

\emailAdd{roncala@nikhef.nl}
\emailAdd{kpetraki@nikhef.nl}

\abstract{
If dark matter (DM) couples to a force carrier that is much lighter than itself, then it may form bound states in the early universe and inside haloes. While bound-state formation via vector emission is known to be efficient and have a variety of phenomenological implications, the capture via scalar emission typically requires larger couplings and is relevant to more limited parameter space, due to cancellations in the radiative amplitude. However, this result takes into account only the trilinear DM-DM-mediator coupling. Theories with scalar mediators include also a scalar potential, whose couplings may participate in the radiative transitions. We compute the contributions of these couplings to the radiative capture, and determine the parameter space in which they are important. 
}

\arxivnumber{1808.04854} 

\begin{document}
\maketitle

\section{Introduction \label{Sec:Intro}}

In a variety of theories, dark matter (DM) is hypothesised to couple to scalar mediators. Such a coupling may determine the DM density via thermal freeze-out, and/or connect the dark sector to the Standard Model (SM) particles. It is a particularly compelling possibility in view of the discovery of the Higgs boson, which may itself be the mediator, or provide a portal to DM via an extended scalar sector. If DM possesses a sizeable coupling to the SM Higgs, then it is constrained by current experiments to be significantly heavier than the Higgs. On the other hand, if DM couples to a non-SM scalar mediator that is much lighter than itself, then this coupling may result in significant DM self-scattering inside haloes that can affect the galactic structure and bring theoretical predictions in better agreement with observations~\cite{Tulin:2017ara}. Even outside the self-interacting DM regime, the coupling of DM to a lighter non-SM scalar mediator is a generic possibility within dark sector models.

In scenarios where DM couples directly to a light force carrier, non-perturbative effects associated with the long-range nature of the interaction impact the DM phenomenology. It is well known that the Sommerfeld effect~\cite{Sommerfeld:1931,Sakharov:1948yq} can influence the DM annihilation and self-scattering rates. This can be the case either if the mediator is a non-SM scalar or the SM Higgs~\cite{Harz:2017dlj}. More recently, it has been realised that the cosmological and astrophysical formation of DM bound states is a generic implication of theories with light mediators. The formation and subsequent decay of unstable bound states can deplete the DM density~\cite{vonHarling:2014kha}, and contribute to the DM indirect detection signals~\cite{Pospelov:2008jd,An:2016gad,Asadi:2016ybp,Petraki:2016cnz,Cirelli:2016rnw,Kouvaris:2016ltf,Baldes:2017gzw,Baldes:2017gzu,MarchRussell:2008tu,An:2016kie}. The formation of stable bound states may quell the DM self-scattering inside haloes~\cite{Petraki:2014uza}, and give rise to novel radiative~\cite{Pearce:2015zca,Cline:2014eaa,Pearce:2013ola,Detmold:2014qqa} and direct detection signatures~\cite{Laha:2013gva,Butcher:2016hic}.  
Attractive interactions may result in the formation of large bound states~\cite{Krnjaic:2014xza,Wise:2014jva,Wise:2014ola} or non-topological solitons~\cite{Coleman:1985ki,Kusenko:1997ad,Kusenko:1997zq,Kusenko:1997si}.

While the importance of bound-state formation (BSF) for DM is now well established for weakly coupled theories with vector mediators~\cite{Pospelov:2008jd,Shepherd:2009sa,vonHarling:2014kha,Petraki:2014uza,Cline:2014eaa,Pearce:2015zca,Petraki:2015hla,Asadi:2016ybp,An:2016gad,Petraki:2016cnz,Cirelli:2016rnw,Kouvaris:2016ltf,Kim:2016kxt,Kim:2016zyy,Baldes:2017gzw,Baldes:2017gzu,Harz:2018csl,Geller:2018biy,Biondini:2018pwp,Biondini:2017ufr,Kaplan:2009de,CyrRacine:2012fz}, and of course in confining DM theories~\cite{Lonsdale:2014wwa,Boddy:2014yra,Kribs:2016cew,Lonsdale:2017mzg,Lonsdale:2018xwd}, in models with scalar mediators bound-state effects appear to be either less severe, or relevant to a limited parameter space. In such models, the radiative capture into bound states suffers from two cancellations:  (i) The lowest order $s$-wave contribution vanishes due to the orthogonality of the incoming and outgoing wavefunctions, thus abdicating the leading order to the $p$-wave. (ii) The leading order $p$-wave terms cancel for particle-antiparticle or identical-particle pairs, yielding to $s$- and $d$-wave contributions that are suppressed by higher orders in the coupling with respect to the capture via vector emission and possibly to the annihilation cross-section~\cite{Wise:2014jva,Petraki:2015hla,Petraki:2016cnz}. Even so, BSF has some important implications. Since it contains an $s$-wave component, it results in CMB constraints for fermionic DM, whose direct annihilation is $p$-wave and thus unconstrained by indirect probes~\cite{An:2016kie}. Moreover, asymmetric DM coupled to a light scalar may form stable multiparticle bound states, provided that the coupling is large enough to overcome the two-particle capture bottleneck~\cite{Wise:2014jva,Wise:2014ola,Gresham:2017cvl} (or BSF is facilitated by a parametric resonance~\cite{Braaten:2018xuw}).

The above results take into account only the trilinear DM-DM-mediator coupling. This coupling is responsible for the long-range interaction between the DM particles that determines the scattering and bound state wavefunctions. The same coupling contributes to the radiative part of the capture process. However, there may be also other couplings that contribute to this piece. Since the radiative vertex enters the overlap integral between the initial and final states, its strength and momentum dependence are critical in determining the efficacy of BSF. It is thus important to consider all relevant contributions, particularly if they are generic within a theory. Theories with scalar mediators include a scalar potential whose couplings can contribute to the radiative part of the capture process. In this work, we investigate the impact of the various couplings in the scalar potential to the radiative BSF in theories with scalar mediators.

The paper is organised as follows. In \cref{Sec:BSF_general},  we introduce the interaction Lagrangians, review the computation of radiative transitions, and summarise the past results on BSF with scalar emission via the trilinear DM-DM-mediator coupling. In \cref{Sec:BSF_SC}, we compute the contributions to the radiative BSF from other scalar couplings. We consider both scalar and fermionic DM, and compute BSF via one and two scalar emission (BSF1 and BSF2 respectively). We discuss the features of the resulting  cross-sections and compare them with the past results. We conclude in \cref{Sec:Conc}, with a discussion of their potential implications. Various technical computations are included in the appendices. For easy reference, in \cref{TableNotation} we summarise the notation used throughout the paper.

\vfill
\setlength{\tabcolsep}{1.5em} 
{\renewcommand{\arraystretch}{1.65}
\begin{table}[h]
\centering
\begin{tabular}{|c|c|}
\hline
\textbf{Description}                          & \textbf{Symbol}						\\ 
\hline \hline 
Interacting DM Particles                      &  $X_1, X_2$							\\ 
\hline 
Mass of the DM interacting particles          &  $m_1,m_2$                     	\\ 
\hline
Total mass of the DM interacting particles    &  $M=m_1+m_2$                       	\\ 
\hline
Reduced mass of the DM interacting particles  &  $\mu=\dfrac{m_1m_2}{m_1+m_2}$     	\\[5pt] 
\hline
Mass ratios									  &  $\eta_{1,2}=\dfrac{m_{1,2}}{m_1+m_2}$ \\[5pt]
\hline
Scalar force mediator                         &  $\varphi$                         	\\ 
\hline
Mass of the scalar force mediator             &  $m_\varphi$						\\ 
\hline  \hline
Dimensionless coupling constants              &  $g_1,g_2$							\\ 
\hline 
Fermionic dark fine structure                 &  $\alpha_f=g_1g_2/(4\pi)$			\\ 
\hline 
Scalar dark fine structure                    &  $\alpha_s=g_1g_2/(16\pi)$           \\ 
\hline 
Bohr momentum                                 &  $\kappa=\mu \alpha$				\\ 
\hline 
Relative velocity of particles in the scattering state  
&  ${\bf \vrel }$     \\ 
\hline 
\parbox[c]{7cm}{
\centering
Momentum of particles in the \\ scattering state in the CM frame}     
&  ${\bf k = \mu \vrel }$     \\[5pt]   
\hline
\multirow{2}{*}{  \parbox[c]{7cm}{\centering
Dimensionless parameters that \\ determine the wavefunctions}}                   
&  $\zeta \equiv \alpha/\vrel $              	\\[5pt]
\cline{2-2} 
&  $\xi \equiv \dfrac{\mu \alpha}{0.84 m_\varphi}$  \\[5pt]
\hline 
Binding energy of $n\ell m$ bound state       
&  $\epsilon_{n\ell}$        \\ 
in the Hulthen approximation       
&  $\epsilon_{n\ell}=\dfrac{\mu \alpha^2}{2 n^2} \left(1-\dfrac{n^2}{\xi}\right)^2$  \\[5pt] 
\hline 
Kinetic energy of scattering state in the CM frame     
&  $\epsilon_{{\bf k}}=\dfrac{{\bf k^2}}{2\mu}=\dfrac{\mu \vrel ^2}{2}$ \\[5pt]
\hline
\end{tabular}
\caption{Notation.}
\label{TableNotation}
\end{table}

\clearpage
\section{Bound-state formation via emission of scalar mediators \label{Sec:BSF_general}}

\subsection{Preliminaries} \label{sec:BSF_general_preliminaries}

We consider two particles $X_1$ and $X_2$ with masses $m_1$ and $m_2$ respectively, that interact via a light scalar mediator $\varphi$ of mass $m_\varphi$. We shall allow $X_1$ and $X_2$ to be real or complex scalars, or Dirac fermions. The relevant interactions are described by the following Lagrangians,
\begin{equation}
\label{eq:L_Scalars_Real}
\begin{split}
\mathcal{L}_{\Re,sc} 
&= \frac{1}{2} \partial_\mu X_1  \partial^\mu X_1 + \frac{1}{2} \partial_\mu X_2  \partial^\mu X_2 + \frac{1}{2} \partial_\mu \varphi\partial^\mu \varphi 
- \frac{1}{2} m_1^2 X_1^2 - \frac{1}{2} m_2^2 X_2^2 - \frac{1}{2} m_\varphi^2 \varphi^2
\\
&- \frac{1}{2} g_1 m_1 \varphi X_1^2 - \frac{1}{2} g_2 m_2 \varphi X_2^2
-\frac{\lambda_{1\varphi}}{4} X_1^2 \varphi^2
-\frac{\lambda_{2\varphi}}{4} X_2^2 \varphi^2
-\frac{\rho_\varphi}{3!} \varphi^3
-\frac{\lambda_\varphi}{4!} \varphi^4
\\
&- \frac{\lambda_1}{4!} X_1^4 - \frac{\lambda_2}{4!} X_2^4  - \frac{\lambda_{12}}{4} X_1^2 X_2^2 \,,
\end{split}
\end{equation}
\begin{equation}
\label{eq:L_Scalars_Complex}
\begin{split}
{\cal L}_{\Im,sc} 
&= \partial_\mu X_1^\dagger  \partial^\mu X_1+ \partial_\mu X_2^\dagger  \partial^\mu X_2+ \frac{1}{2}\partial_\mu \varphi \partial^\mu \varphi 
- m_1^2 |X_1|^2 - m_2^2 |X_2|^2 - \frac{1}{2} m_\varphi^2 \varphi^2
\\
&- g_1 m_1 \varphi |X_1|^2 - g_2 m_2 \varphi |X_2|^2
-\frac{\lambda_{1\varphi}}{2} |X_1|^2 \varphi^2
-\frac{\lambda_{2\varphi}}{2} |X_2|^2 \varphi^2
-\frac{\rho_\varphi}{3!} \varphi^3
-\frac{\lambda_\varphi}{4!} \varphi^4
\\
&- \frac{\lambda_1}{2} |X_1|^4 - \frac{\lambda_2}{2} |X_2|^2  - \lambda_{12} |X_1|^2 |X_2|^2 \,,
\end{split}
\end{equation}
and
\begin{equation}
\label{eq:L_Fermions}
\begin{split}
{\cal L}_f &= 
 \bar{X}_1 i\slashed{\partial} X_1  
+\bar{X}_2 i\slashed{\partial} X_2+ \frac{1}{2}\partial_\mu \varphi  \partial^\mu \varphi
- m_1 \bar{X}_1 X_1 - m_2  \bar{X}_2 X_2 - \frac{1}{2} m_\varphi^2 \varphi^2 
\\
&- g_1 \varphi \bar{X}_1 X_1 - g_2 \varphi \bar{X}_2 X_2 
-\frac{\rho_\varphi}{3!} \varphi^3
-\frac{\lambda_\varphi}{4!} \varphi^4 \,.
\end{split}
\end{equation}
Note that, since we are interested in the application of our results to DM, we shall assume that the interacting particles $X_1,X_2$ carry a $\mathbb{Z}_2$ symmetry.

For later convenience, we define the total and the reduced mass of the two interacting particles
\begin{equation}
M \equiv m_1 + m_2, \qquad \mu \equiv \frac{m_1 m_2}{m_1+m_2} \,,
\label{eq:masses}
\end{equation}
and the dimensionless factors
\begin{equation}
\eta_1  \equiv \frac{m_1}{m_1+m_2}, \qquad \eta_2 \equiv \frac{m_2}{m_1+m_2} \,.
\label{eq:etas}
\end{equation}

In the non-relativistic regime, the interaction between $X_1$ and $X_2$ is described to leading order by a static Yukawa potential that arises from the resummation of the one-boson-exchange diagrams,
\begin{equation}
\label{eq:Yukawa}
V_Y({\bf r})=-\frac{\alpha}{r}e^{-m_\varphi r},
\end{equation}
with $\alpha = \alpha_{sc}$ or $\alpha = \alpha_f$, depending on whether the interacting particles are scalars or fermions, where
\begin{equation}
\alpha_{sc} \equiv \frac{g_1 g_2}{16\pi} 
\qquad \text{and} \qquad
\alpha_{f} \equiv \frac{g_1 g_2}{4\pi} \,.
\label{eq:alpha_scANDf}
\end{equation}
We derive the Yukawa potential and $\alpha_{sc}, \alpha_f$ in \cref{App:potential}. 
The long-range interaction between $X_1$ and $X_2$ described by the potential \eqref{eq:Yukawa} distorts the wavefunction of the scattering (unbound) states -- a phenomenon known as the Sommerfeld effect~\cite{Sommerfeld:1931,Sakharov:1948yq} -- and gives rise to bound states. Bound states exist if the mediator is sufficiently light. For the ground state to exist,
\begin{equation}
\mu \alpha / m_\varphi > 0.84 \,,
\label{eq:BoundStateExistence}
\end{equation}
while stronger conditions apply for excited states~\cite{Petraki:2016cnz}. The condition \eqref{eq:BoundStateExistence} also roughly marks the regime where the Sommerfeld effect is significant.

The capture into bound states necessitates the dissipation of the binding energy and the kinetic energy of the relative motion of the $X_1 X_2$ pair, which may occur radiatively. In \cref{sec:BSF_general_RadiativeTransAmpl}, we review the computation of radiative BSF amplitudes. 
The capture via emission of one scalar mediator, $X_1 + X_2 \to {\cal B}(X_1 X_2) + \varphi$, has been previously considered in Refs.~\cite{Wise:2014jva,Petraki:2015hla,An:2016kie,Petraki:2016cnz}, where only the trilinear $\varphi X_j^2$ couplings were taken into account. We review the main results in \cref{sec:BSF_general_1phi_TC}. As we shall see, for a particle-antiparticle pair or a pair of identical particles, the dipole contribution -- which is the leading order term for $X_1, X_2$ with different masses and couplings -- vanishes identically.

\subsection{Radiative transition amplitude \label{sec:BSF_general_RadiativeTransAmpl}}

We consider the radiative transitions
\begin{equation}
X_1 (k_1) + X_2 (k_2) \to 
X_1 (p_1) + X_2 (p_2) + \text{radiation} \,,
\label{eq:RadiativeTransition}
\end{equation}
where the parentheses denote the 4-momenta of the incoming and outgoing $X_1$ and $X_2$ fields.
We will be interested in particular in the case where the incoming  $X_1,X_2$ particles form a scattering state, while the outgoing $X_1,X_2$ are captured into a bound state. In order to separate the motion of the CM from the relative motion, we make the following transformation in the momenta~\cite{Itzykson:1980rh,Petraki:2015hla}
\begin{align}
k_1 \equiv \eta_1 K + q, \qquad k_2 \equiv \eta_2 K - q \,, 
\label{eq:MomentumTrans_k} \\
p_1 \equiv \eta_1 P + p, \qquad p_2 \equiv \eta_2 P - p \,,
\label{eq:MomentumTrans_p}
\end{align}
where $\eta_{1,2}$ are defined in \cref{eq:etas}. 

In the presence of a long-range interaction, the relative motion of $X_1, X_2$ is not well approximated by a plane wave. It is described more generally by wavefunctions,
which in momentum space we shall denote as
$\tilde{\phi}_{\bf k} ({\bf q})$ and $\tilde{\psi}_{n\ell m} ({\bf p})$ for the scattering and the bound states respectively. The wavefunctions obey the Schr\"odinger equation with the potential \eqref{eq:Yukawa}. 
The continuous spectrum is characterised by the momentum ${\bf k} = \mu {\bf v}_{\rm rel}$, which is the expectation value of ${\bf q}$ and parametrises the energy of the relative motion in the scattering states, $\epsilon_{\bf k} = {\bf k}^2/(2\mu) = \mu \vrel^2/2$. The bound states are characterised by the standard discrete principal and angular momentum quantum numbers $\{n\ell m\}$, which determine the expectation value of ${\bf p}$ and the binding energy $\epsilon_{n\ell}$. As is well known, for a Coulomb potential, $\epsilon_{n\ell} = \kappa^2/(2n^2\mu) = \mu \alpha^2/(2n^2)$, where $\kappa \equiv \mu \alpha$ is the Bohr momentum; a non-negligible mediator mass suppresses $\epsilon_{n\ell}$ and introduces a dependence on $\ell$. We review the wavefunctions in \cref{App:wf} (see Ref.~\cite{Petraki:2016cnz} for a more detailed discussion). For the purpose of evaluating the leading order contributions to the transition amplitude, we shall keep in mind that the wavefunctions impose   
$|{\bf q}| \sim |{\bf k}| = \mu \vrel$ and 
$|{\bf p}| \sim \kappa = \mu \alpha$.

In the non-relativistic regime, the total 4-momenta of the scattering and the bound states are
\begin{align}
K &\simeq \left(M + \frac{{\bf K}^2}{2M} + \epsilon_{\bf k}, \ {\bf K} \right) \,,
\label{eq:K} \\
P &\simeq \left(M + \frac{{\bf P}^2}{2M} - \epsilon_{n\ell}, \ {\bf P} \right) \,.
\label{eq:P}
\end{align}
We will work in the CM frame, ${\bf K=0}$. 
Then, taking into account that $\epsilon_{\bf k}, \epsilon_{n\ell} \ll M$ (or equivalently $\alpha, \vrel \ll 1$), the total energy available to be dissipated is 
\begin{equation}
\omega \simeq \epsilon_{\bf k} + \epsilon_{n\ell} \,.
\label{eq:omega}
\end{equation}
Evidently, the bound state acquires momentum $|{\bf P}| \sim \omega$.

The full amplitude for the radiative capture into a bound state depends on the overlap of the initial (scattering) and final (bound) state wavefunctions and the radiative vertex. The diagrammatic representation for transitions with emission of one or two scalars is shown in \cref{fig:BSF_FullAmplitude}. In the instantaneous and non-relativistic approximations, the amplitude is~\cite{Petraki:2015hla}
\begin{equation}
\mathcal{M}_{{\bf k}\rightarrow \{n\ell m\}}  \simeq 
\frac{1}{\sqrt{2\mu}} \int
\frac{d^3p}{(2\pi)^3}
\frac{d^3q}{(2\pi)^3}
\left[1-\frac{\bf { p^2}+q^2}{4\mu^2}\left( 1-\frac{3\mu}{M} \right)  \right]
\tilde{\psi}^*_{n\ell m}({\bf p}) 
\tilde{\phi}_{\bf k}({\bf q})
\mathcal{A}_T({\bf q,p}) \,,
\label{eq:MBSF}
\end{equation}
where ${\cal A}_T ({\bf q,p})$ is the radiative amplitude for the (off-shell) transition \eqref{eq:RadiativeTransition}, under the transformations of \cref{eq:MomentumTrans_k,eq:MomentumTrans_p}. The fully connected diagrams contributing to ${\cal A}_T$ can be evaluated at leading order by setting the incoming and outgoing $X_1,X_2$ on-shell. For any non-fully-connected diagrams contributing to ${\cal A}_T$, the virtuality of $X_1$ and $X_2$ has to be integrated out. This can be done starting from the off-shell amplitude as described in \cite{Petraki:2015hla} (see also \cite[section 2.3]{Harz:2018csl} for a brief summary), or by adopting an  effective field theory approach~\cite{Pineda:1997bj,Brambilla:1999xf,Beneke:1999zr,Manohar:1999xd,Manohar:2000kr,Pineda:2001ra,Brambilla:2004jw,Pineda:2011aw,Hoang:2011gy,Braaten:2017kci,Braaten:2017gpq,Braaten:2017dwq}.\footnote{For a comparison of quantum and classical approaches, see~\cite{Belotsky:2015osa}.}
The dominant contributions to the capture with emission of a vector or scalar mediator via the trilinear coupling arise from non-fully-connected diagrams~\cite{Petraki:2015hla,Petraki:2016cnz}. (However, in non-Abelian theories, a leading order contribution to capture via gluon emission arises also from a fully connected diagram~\cite{Asadi:2016ybp}.)  
In the computations of this paper in \cref{Sec:BSF_SC}, we will consider only fully connected diagrams.

In \cref{eq:MBSF}, the factor inside the square brackets includes the leading order corrections in ${\bf {\bf p^2}, q^2}$ arising from the relativistic normalisation of states~\cite{Petraki:2015hla}. Upon the convolution with the wavefunctions and integration over ${\bf p}$ and ${\bf q}$, these terms amount to corrections in $\alpha^2$ and $\vrel^2$. They become important when the leading order contribution from ${\cal A}_T$ alone cancels, as is the case for the capture of particle-antiparticle or identical-particle pairs with emission of a scalar mediator via the trilinear coupling.

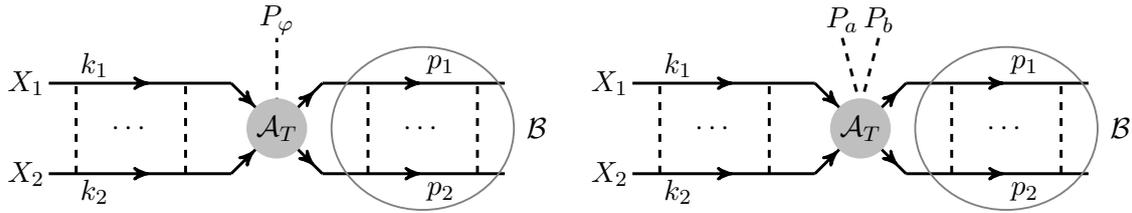
\begin{figure}[t]
\centering
\begin{tikzpicture}[line width=1.1pt, scale=1.2]
\begin{scope}[shift={(-3.2,0)}]
\node at (-2.0, 1.2) {$k_1$};
\node at (-2.0, -0.2) {$k_2$};
\node at (-2.75, 1) {$X_1$};
\node at (-2.75, 0) {$X_2$};
\node at (1.8, 1.2) {$p_1$};
\node at (1.8, -0.2) {$p_2$};
\draw[dashed]    (-2.2,0) --  (-2.2,1);
\node at (-1.6,0.5) {$\cdots$};
\draw[dashed]    (-1,0) -- (-1,1);
\draw[fermion] (-2.5,1) -- (-0.5,1);
\draw[fermion] (-0.5,1) -- (-0.05,0.55);
\draw[fermion] (-0.5,0) -- (-0.05,0.45);
\draw[fermion] (-2.5,0) -- (-0.5,0);
\draw[dashed]    (2.2,0) -- (2.2,1);
\node at (1.6,0.5) {$\cdots$};
\draw[dashed]    (1,0) -- (1,1);
\draw[fermionbar] (2.5,1) -- (0.5,1);
\draw[fermionbar] (0.5,1) -- (0.1,0.6);
\draw[fermionbar] (0.5,0) -- (0.1,0.4);
\draw[fermionbar] (2.5,0) -- (0.5,0);
\draw[line width=0.7pt,gray] (1.6,0.5) ellipse (1cm and 0.9cm);
\node at (2.85,0.5) {${\cal B}$};
\draw[dashed] (0,0.75) -- (0,1.5);
\node at (0,1.7) {$P_\varphi$};
\filldraw[lightgray]  (0,0.5) circle (9pt);
\node at (0,0.5) {${\cal A}_T$};
\end{scope}
\begin{scope}[shift={(3.2,0)}]
\node at (-2.0, 1.2) {$k_1$};
\node at (-2.0, -0.2) {$k_2$};
\node at (-2.75, 1) {$X_1$};
\node at (-2.75, 0) {$X_2$};
\node at (1.8, 1.2) {$p_1$};
\node at (1.8, -0.2) {$p_2$};
\draw[dashed]    (-2.2,0) --  (-2.2,1);
\node at (-1.6,0.5) {$\cdots$};
\draw[dashed]    (-1,0) -- (-1,1);
\draw[fermion] (-2.5,1) -- (-0.5,1);
\draw[fermion] (-0.5,1) -- (-0.05,0.55);
\draw[fermion] (-0.5,0) -- (-0.05,0.45);
\draw[fermion] (-2.5,0) -- (-0.5,0);
\draw[dashed]    (2.2,0) -- (2.2,1);
\node at (1.6,0.5) {$\cdots$};
\draw[dashed]    (1,0) -- (1,1);
\draw[fermionbar] (2.5,1) -- (0.5,1);
\draw[fermionbar] (0.5,1) -- (0.1,0.6);
\draw[fermionbar] (0.5,0) -- (0.1,0.4);
\draw[fermionbar] (2.5,0) -- (0.5,0);
\draw[line width=0.7pt,gray] (1.6,0.5) ellipse (1cm and 0.9cm);
\node at (2.85,0.5) {${\cal B}$};
\draw[dashed] (-0.2,1.5) -- (0,0.75);\node at (-0.2,1.7) {$P_a$};
\draw[dashed] (0.2,1.5) -- (0,0.75);\node at (0.2,1.7) {$P_b$};
\filldraw[lightgray]  (0,0.5) circle (9pt);
\node at (0,0.5) {${\cal A}_T$};
\end{scope}
\end{tikzpicture}
\caption{\label{fig:BSF_FullAmplitude} 
The amplitude for the radiative capture into bound states via emission of one or two scalars consists of the initial and final state wavefunctions, and the perturbative radiative amplitude ${\cal M}_T$ that includes the radiative vertices.}
\end{figure}

\subsection{Capture with scalar emission via the trilinear coupling  \label{sec:BSF_general_1phi_TC}}

\begin{figure}[t]
\centering
\begin{tikzpicture}[line width=1.1pt, scale=1.5]
\begin{scope}[shift={(-1.5,0)}]
\draw[fermionnoarrow] (-1,1) -- (1,1);
\draw[fermionnoarrow] (-1,0) -- (1,0);
\draw[scalarnoarrow]   (0,1) -- (0,1.9);
\node at (-1.5,1) {$X_1$};
\node at (-1.5,0) {$X_2$};
\node at (   0,2.1) {$\varphi$};
\draw[->] (-0.9, 1.1) -- (-0.5, 1.1);
\draw[->] ( 0.5, 1.1) -- ( 0.9, 1.1);
\draw[->] (-0.9,-0.1) -- (-0.5,-0.1);
\draw[->] ( 0.5,-0.1) -- ( 0.9,-0.1);
\draw[->] ( 0.1, 1.4) -- ( 0.1, 1.85);
\node at (-0.8, 1.3) {$\eta_1 K + q$};
\node at ( 0.8, 1.3) {$\eta_1 P + p$};
\node at (-0.8,-0.3) {$\eta_2 K - q$};
\node at ( 0.8,-0.3) {$\eta_2 P - p$};
\node at ( 0.4, 1.7) {$P_\varphi$};
\node at ( 0, 0.8) {$g_1 m_1$};
\end{scope}
\begin{scope}[shift={(1.5,0)}]
\draw[fermionnoarrow] (-1,1) -- (1,1);
\draw[fermionnoarrow] (-1,0) -- (1,0);
\draw[scalarnoarrow]   (0,0) -- (0,-0.9);
\draw[->] ( 0.1,-0.4) -- ( 0.1,-0.85);
\node at ( 0.4,-0.7) {$P_\varphi$};
\node at ( 0,0.15) {$g_2 m_2$};
\end{scope}
\end{tikzpicture}
\caption{\label{fig:BSF1_trilinear} 
The contribution of the trilinear DM-DM-mediator couplings to the radiative part of bound-state formation via emission of one scalar mediator.}
\end{figure}
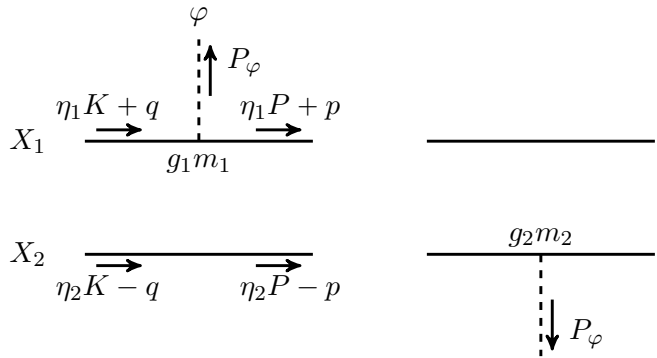

The leading contributions to capture with emission of a scalar via the trilinear coupling only are shown in \cref{fig:BSF1_trilinear}. Starting from \cref{eq:MBSF}, and neglecting the correction arising from the normalisation of states, these diagrams yield\footnote{
We use the superscript $\TC$ to denote contributions that arise from the trilinear coupling alone, and in \cref{Sec:BSF_SC} we will use the superscript $\SC$ to denote contributions in which other scalar couplings participate.
}~\cite{Petraki:2015hla,Petraki:2016cnz}
\begin{equation}
{\cal M}_{{\bf k} \to n\ell m}^\TC 
\simeq -M\sqrt{2\mu}
\int d^3 r \ \psi_{n\ell m}^* ({\bf r}) \ \phi_{\bf k} ({\bf r})
\left(
g_1 \, e^{-i \eta_2 {\bf P}_\varphi \cdot {\bf r}}  + 
g_2 \, e^{i \eta_1 {\bf P}_\varphi \cdot {\bf r}}  
\right) \,,
\label{eq:MBSF1_nondeg}
\end{equation}
where ${\bf P}_\varphi$ is the momentum of the emitted boson, with 
$
{\bf P}_\varphi^2 + m_\varphi^2 = \omega^2
$
and $\omega$ given in \cref{eq:omega}. The wavefunctions imply that the integrand is significant in the region $r \lesssim 1/\max(\mu \alpha/n,\mu \vrel) \ll 1/\omega$, therefore we may evaluate \cref{eq:MBSF1_nondeg} by expanding in ${\bf P}_\varphi \cdot {\bf r} \sim \max(\alpha/n,\vrel)$ (cf.~ref.~\cite[appendix~B]{Petraki:2016cnz}). Clearly, the zeroth order terms vanish due to the orthogonality of the wavefunctions.

For a pair of particles with different masses and/or couplings to the scalar mediator, the leading order contributions arise from the ${\bf P}_\varphi \cdot {\bf r}$ terms in the expansion of \cref{eq:MBSF1_nondeg}. The cross-section for capture into the ground state $\{n\ell m\} = \{100\}$ is~\cite{Petraki:2015hla,Petraki:2016cnz}
\begin{equation}
\sigma_{\BSFone}^\TC \vrel \simeq 
\left[ \frac{(g_1\eta_2-g_2\eta_1)^2}{16\pi\alpha} \right]
\, \frac{\pi \alpha^2}{\mu^2}
\ S_{\BSFone}^\TC \,,
\label{eq:sigmaBSF1_nondeg}
\end{equation}
where $S_{\BSFone}^\TC$ depends on the dimensionless parameters  $\alpha/\vrel$ and $\mu\alpha/m_\varphi$. In the Coulomb approximation, which is valid at $m_\varphi \lesssim \mu \vrel$, $S_{\BSFone}^\TC$ depends only on the parameter $\zeta \equiv \alpha / \vrel$ as follows~\cite{Petraki:2015hla,Petraki:2016cnz}
\begin{equation}
S_{\BSFone}^\TC (\zeta)= 
\left( \frac{2\pi \zeta}{1-e^{-2\pi \zeta}} \right) \times
\frac{2^6}{3} 
\left(\frac{\zeta^2}{1+\zeta^2}\right)^2
e^{-4\zeta {\rm arccot} \, \zeta} \,.
\label{eq:SBSF1_nondeg}
\end{equation}
For the more general case that includes the effect of the mediator mass, we refer to~\cite{Petraki:2016cnz}. We note that \cref{eq:sigmaBSF1_nondeg,eq:SBSF1_nondeg} correspond to dipole emission ($\ell_S=1$ mode of the scattering state wavefunction).  In \cref{eq:sigmaBSF1_nondeg}, the factor in the square brackets reduces to 1 for a pair of scalar particles with $g_1=g_2$ and $\eta_1 \gg \eta_2$.  
In \cref{eq:SBSF1_nondeg}, the first factor inside the brackets is responsible for the characteristic $\sigma \vrel~\propto~1/\vrel$ scaling of the Sommerfeld-enhanced processes at low velocities ($\zeta \gtrsim 1$), while the remaining factors tend to a constant.

Evidently, for a particle-antiparticle pair or a pair of identical particles ($g_1=g_2=g$ and $\eta_1=\eta_2=1/2$), the two terms proportional to ${\bf P_\varphi}$ in the expansion of \cref{eq:MBSF1_nondeg} cancel. This cancellation persists even for capture into bound states of non-zero angular momentum ($\ell >0$), as is evident from \cref{eq:MBSF1_nondeg}, and implies that the next order contributions should be considered.   

For a scalar particle-antiparticle pair the next order terms, which include also the leading order correction from the normalisation of states shown in \cref{eq:MBSF}, yield~\cite{Petraki:2016cnz}
\begin{equation}
\sigma_{\BSFone}^\TC \vrel \simeq 
\, \frac{\pi \alpha_{sc}^4}{\mu^2}
\ S_{\BSFone,\mathsmaller{XX^*}}^\TC  \,,
\label{eq:sigmaBSF1_deg}
\end{equation}
where in the Coulomb regime $S_{\BSFone,\mathsmaller{XX^*}}^\TC$ depends on $\zeta = \alpha_{sc} / \vrel$ as follows
\begin{equation}
S_{\BSFone,\mathsmaller{XX^*}}^\TC \, (\zeta)= 
\left( \frac{2\pi \zeta}{1-e^{-2\pi \zeta}} \right) \times
\frac{2^6}{15} 
\frac{\zeta^2(3+2\zeta^2)}{(1+\zeta^2)^2}
\ e^{-4\zeta {\rm arccot} \, \zeta} \,.
\label{eq:SBSF1_Coulomb_deg}
\end{equation}
(For a pair of identical scalars, an extra factor of 2 arises from the symmetrization of the scattering state wavefunction.) The case of fermionic DM was considered in ref.~\cite{Wise:2014jva}. In the Coulomb regime and for $\zeta \gtrsim 1$, the cross-section was found to be
\begin{equation}
\sigma_{\BSFone,\mathsmaller{X\bar{X}}}^\TC \, \vrel \simeq 
\frac{\pi \alpha_f^4}{\mu^2}
\left( \frac{2\pi \alpha_f^{}}{\vrel} \right)
\frac{2^4 e^{-4}}{3^2}\,.
\label{eq:sigmaBSF1_degF}
\end{equation}
Note that both \cref{eq:sigmaBSF1_deg} and \cref{eq:sigmaBSF1_degF} are suppressed by $\alpha^2$ with respect to \cref{eq:sigmaBSF1_nondeg}.

\clearpage
\section{\label{Sec:BSF_SC} The contribution of the scalar potential couplings to radiative capture}

In this section, we investigate how the scalar couplings $\lambda_{1\varphi}$, $\lambda_{2\varphi}$, $\lambda_\varphi$ and $\rho_\varphi$ in \cref{eq:L_Scalars_Real,eq:L_Scalars_Complex,eq:L_Fermions} -- which do not affect the long-range potential between $X_1$ and $X_2$ -- contribute to the radiative capture of $X_1, X_2$ pairs into bound states. In order to exhibit the leading order contribution from all couplings, we consider BSF via one and two scalar emission,
\begin{align}
\text{BSF1:} \quad 
X_1 + X_2 &\to {\cal B}(X_1 X_2) + \varphi \,,
\label{eq:BSFvia1phi_process}
\\
\text{BSF2:} \quad 
X_1 + X_2 &\to {\cal B}(X_1 X_2) + 2\varphi \,.
\label{eq:BSFvia2phi_process}
\end{align}

It is important to note that in most diagrams we will consider, the trilinear couplings $g_j$ also participate in the radiative part of the process (cf.~\cref{fig:BSF1_scalar,fig:BSF2_MediatorTrilinear,fig:BSF2_MediatorQuartic,fig:BSF2_lambdaj}). In fact, it may naively seem that these diagrams are of the same or higher order in $\alpha$ than those giving rise to the cross-sections of \cref{sec:BSF_general_1phi_TC} (even ignoring the additional suppression introduced by the new couplings). However, the momentum transfer along the mediators exchanged in these diagrams and the off-shellness of the interacting particles scale also with $\alpha$, thereby reducing the order of dependence of the diagrams on $\alpha$.\footnote{This is of course also the reason for the emergence of the non-perturbative effects we consider, the Sommerfeld enhancement and the existence of bound states.} 

For convenience, we first define in \cref{sec:BSF_SC_OverlapIntegrals} the wavefunction overlaps integrals that we will use for the BSF cross-sections. We provide some analytic approximations for capture into the ground state, and describe their features that are of course inherited by the BSF cross-sections. Then, in \cref{sec:BSF_SC_1phi_ScalarX,sec:BSF_SC_1phi_FermionX} we compute the contributions to BSF1 from the couplings in the scalar potential, for scalar and fermionic interacting particles, respectively, and discuss in which regimes they are important. We finish with considering BSF2 in \cref{sec:BSF_SC_2phi}.

\subsection{Overlap integrals \label{sec:BSF_SC_OverlapIntegrals}}
We define the overlap integrals
\begin{align}
{\cal V}_{{\bf k},\{n\ell m\}} 
&\equiv (8\pi \kappa)^{1/2} 
\int \frac{d^3 q}{(2\pi)^3} \frac{d^3 p}{(2\pi)^3}
\frac
{\tilde{\phi}_{\bf k} ({\bf q}) \tilde{\psi}_{n\ell m}^* ({\bf p})}
{({\bf q}-{\bf p})^2 + m_\varphi^2} \,,
\label{eq:Overlap_V_def}
\\
{\cal R}_{{\bf k},\{n\ell m\}} 
&\equiv (8 \pi \kappa^5)^{1/2}
\int \frac{d^3 q}{(2\pi)^3} \frac{d^3 p}{(2\pi)^3}
\frac
{\tilde{\phi}_{\bf k} ({\bf q}) \tilde{\psi}_{n\ell m}^* ({\bf p})}
{[({\bf q}-{\bf p})^2 + m_\varphi^2]^2} \,,
\label{eq:Overlap_R_def}
\\
{\cal I}_{{\bf k},\{n\ell m\}} ({\bf \Gamma})
&\equiv 
\int \frac{d^3 p}{(2\pi)^3}
\tilde{\phi}_{\bf k} ({\bf p + \Gamma}) \tilde{\psi}_{n\ell m}^* ({\bf p})
= \int \frac{d^3 p}{(2\pi)^3}
\phi_{\bf k} ({\bf r}) \psi_{n\ell m}^* ({\bf r})
\, e^{-i {\bf \Gamma}\cdot {\bf r}}
\,.
\label{eq:Overlap_I_def}
\end{align} 
The prefactors in \cref{eq:Overlap_V_def,eq:Overlap_R_def} have been chosen such that ${\cal V}_{{\bf k},\{n\ell m\}}$ and ${\cal R}_{{\bf k},\{n\ell m\}}$ are dimensionless, and the definition of ${\cal I}_{{\bf k},\{n\ell m\}}$ follows refs.~\cite{Petraki:2015hla,Petraki:2016cnz}.

The integrals \cref{eq:Overlap_V_def,eq:Overlap_R_def,eq:Overlap_I_def} depend on the two dimensionless parameters
\begin{equation}
\zeta \equiv \frac{\alpha}{\vrel} 
\quad \text{and} \quad 
\xi \equiv \frac{\mu \alpha}{0.84 m_\varphi} \,.
\label{eq:zetaANDxi_re}
\end{equation}
As stated in \cref{sec:BSF_general_preliminaries}, for the Yukawa potential \eqref{eq:Yukawa} the ground state exists if $\xi > 1$~\cite{Petraki:2016cnz}.\footnote{
Note that in ref.~\cite{Petraki:2016cnz}, the parameter $\xi$ was defined as $\mu \alpha/m_\varphi$. The reason we prefer the definition \eqref{eq:zetaANDxi_re} here is to simplify the expressions for the wavefunctions and the binding energies in the Hulthen approximation of the Yukawa potential.}
However, BSF1 is kinematically possible only if 
\begin{equation}
m_\varphi < (\mu/2) [\alpha^2 (1-1/\xi)^2+ \vrel^2] \,,
\label{eq:BSF1_KinematicThresshold}
\end{equation}
where we used the Hulthen approximation for the binding energy (cf.~\cref{App:wf}).  In the regime where BSF is important, $\vrel \lesssim \alpha$~(see e.g.~\cref{sec:BSF_general_1phi_TC}), this condition reduces roughly to $m_\varphi \lesssim \mu \alpha^2 / 2$ or equivalently $\xi \gtrsim \xi_{\min} \simeq 2.4 / \alpha \gg1$, i.e.~it is much stronger than the requirement for the existence of bound states. This in turn ensures that the bound-state wavefunction can be approximated by its Coulomb value.  (For the validity of the Coulomb limit for the overlap integrals and the BSF cross-sections, see below.)

We derive analytical expressions for the ${\cal V}_{{\bf k},\{100\}}$ and ${\cal R}_{{\bf k},\{100\}}$ integrals in \cref{App:OverlapIntegrals}, using an appropriate approximation. The integral \eqref{eq:Overlap_I_def} has been considered in ref.~\cite{Petraki:2016cnz}. For the BSF cross-sections of interest, $|{\bf \Gamma}| \sim |{\bf P}_\varphi| \ll \langle|{\bf p}|\rangle \sim \kappa$, thus ${\cal I}_{{\bf k} \to \{n\ell m\}}$ can be computed by expanding in ${\bf \Gamma}$. Here we shall keep up to first order terms in  $\Gamma/\kappa$. For capture into the ground state,
\begin{align}
{\cal V}_{{\bf k},\{100\}}  &\simeq 
\sqrt{8 S_0(\zeta, \xi)} 
\left(\frac{\zeta^2}{1+\zeta^2} \right) 
\ e^{-2\zeta \, {\rm arccot} \, \zeta} \,,
\label{eq:V_GroundState_re}
\\
{\cal R}_{{\bf k},\{100\}}  &\simeq 
\sqrt{8 S_0(\zeta, \xi)} 
\left(\frac{\zeta^2}{1+\zeta^2} \right)^2 
\ e^{-2\zeta \, {\rm arccot} \, \zeta}  \,,
\label{eq:R_GroundState_re}
\\
{\cal I}_{{\bf k},\{100\}} ({\bf \Gamma}) &\simeq 
\sqrt{\frac{2^8 \pi S_1(\zeta, \xi)}{\kappa^3}} 
\left[\frac{\zeta^5}{(1+\zeta^2)^3} \right]
\ e^{-2\zeta \, {\rm arccot} \, \zeta} 
\ \frac{\Gamma \cos( \theta_{\bf k, \Gamma})}{\kappa} \,,
\label{eq:I_GroundState_re}
\end{align}
where $S_0(\zeta,\xi)$ and $S_1(\zeta,\xi)$ are the Sommerfeld factors for $s$- and $p$-wave annihilation respectively. In the Hulthen approximation, 
\begin{equation}
S_0(\zeta,\xi) =
\frac{2\pi \zeta \sinh (\pi \xi/\zeta)}
{\cosh(\pi\xi/\zeta) - \cosh [(\pi\xi/\zeta)\sqrt{1-4\zeta^2/\xi} ]} \,,
\label{eq:Hulthen_S0_re}
\end{equation}
which in the Coulomb limit reduces to $S_0^C (\zeta) = 2\pi\zeta/(1-e^{-2\pi\zeta})$. While there is no analytical approximation for $S_1(\zeta, \xi)$ for finite $\xi$, in the Coulomb limit $\xi \to \infty$ it becomes $S_1^C(\zeta)=(1+\zeta^2)S_0^C(\zeta)$.  The Coulomb limit remains a good approximation as long the average momentum transfer between the interacting particles is larger than the mediator mass,
\begin{equation}
m_\varphi \lesssim \mu \vrel \,,
\label{eq:CoulombApprox_Validity}
\end{equation}
or equivalently $\xi > \zeta$ (see e.g.~\cite{Petraki:2016cnz}). 

Outside this range, i.e.~at low velocities $\vrel \lesssim m_\varphi/\mu$, both $S_0$ and $S_1$ exhibit parametric resonances at discrete values of $\xi$ that correspond to the thresholds for the existence of $\ell=0$ and $\ell=1$ bound states respectively. For $S_0$ in the Hulthen approximation, these are  $\xi = n^2$ with $n\in$~integers. 
At non-resonant parametric points, $S_0$ and $S_1$ follow the Coulomb approximation as long as \eqref{eq:CoulombApprox_Validity} is satisfied, but saturate to their respective Coulomb values at $\vrel \approx m_\varphi/\mu$ as the velocity decreases. 
In contrast, if close to a resonant parametric point,  $S_0$ and $S_1$ grow faster than $1/\vrel$ and $1/\vrel^3$ respectively at $\vrel \lesssim m_\varphi/\mu$ (in particular $S_0~\propto~\zeta^2~\propto~1/\vrel^2$), to eventually saturate to a constant value at a lower velocity that depends on the proximity to the resonant point. For $S_0$ in the Hulthen approximation, the saturated value is $S_0 (\zeta, \xi) \simeq \pi^2 \xi / \sin^2 \sqrt{\pi^2\xi}$ [cf.~\cref{eq:Hulthen_S0_re}]. 

This behaviour implies that an $s$-wave cross-section times relative velocity saturates to a constant value at low velocities, while a $p$-wave recovers the velocity suppression that appears in perturbative cross-sections, $\sigma^{p-\rm wave} \vrel~\propto~\vrel^2$, albeit is enhanced with respect to its value if the Sommerfeld effect were neglected. This point will be important in our discussion of the new contributions to BSF that we compute in the following.

\subsection{Capture via emission of one scalar mediator $\varphi$,  scalar $X_1$ and $X_2$  \label{sec:BSF_SC_1phi_ScalarX}}

\subsubsection{Amplitude}

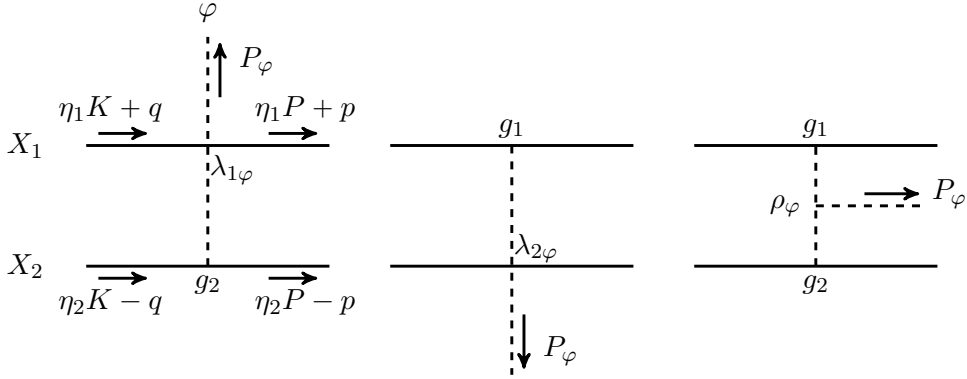
\begin{figure}[t]
\centering
\begin{tikzpicture}[line width=1.1pt, scale=1.6]
\begin{scope}[shift={(-2.5,0)}]
\draw[fermionnoarrow] (-1,1) -- (1,1);
\draw[fermionnoarrow] (-1,0) -- (1,0);
\draw[scalarnoarrow]   (0,0) -- (0,1.9);
\node at (-1.5,1) {$X_1$};
\node at (-1.5,0) {$X_2$};
\node at (   0,2.1) {$\varphi$};
\draw[->] (-0.9, 1.1) -- (-0.5, 1.1);
\draw[->] ( 0.5, 1.1) -- ( 0.9, 1.1);
\draw[->] (-0.9,-0.1) -- (-0.5,-0.1);
\draw[->] ( 0.5,-0.1) -- ( 0.9,-0.1);
\draw[->] ( 0.1, 1.4) -- ( 0.1, 1.85);
\node at (-0.8, 1.3) {$\eta_1 K + q$};
\node at ( 0.8, 1.3) {$\eta_1 P + p$};
\node at (-0.8,-0.3) {$\eta_2 K - q$};
\node at ( 0.8,-0.3) {$\eta_2 P - p$};
\node at ( 0.4, 1.7) {$P_\varphi$};
\node at ( 0.2, 0.8) {$\lambda_{1\varphi}$};
\node at ( 0,-0.15) {$g_2$};
\end{scope}
\begin{scope}[shift={(0,0)}]
\draw[fermionnoarrow] (-1,1) -- (1,1);
\draw[fermionnoarrow] (-1,0) -- (1,0);
\draw[scalarnoarrow]   (0,1) -- (0,-0.9);
\draw[->] ( 0.1,-0.4) -- ( 0.1,-0.85);
\node at ( 0.4,-0.7) {$P_\varphi$};
\node at ( 0, 1.14) {$g_1$};
\node at ( 0.2,0.15) {$\lambda_{2\varphi}$};
\end{scope}
\begin{scope}[shift={(2.5,0)}]
\draw[fermionnoarrow] (-1,1) -- (1,1);
\draw[fermionnoarrow] (-1,0) -- (1,0);
\draw[scalarnoarrow]   (0,0) -- (0,1);
\draw[scalarnoarrow]   (0,0.5) -- (0.9,0.5);
\draw[->] (0.4,0.6) -- (0.85,0.6);
\node at (1.1,0.6) {$P_\varphi$};
\node at ( 0, 1.14) {$g_1$};
\node at ( 0,-0.15) {$g_2$};
\node at (-0.25,0.5)  {$\rho_\varphi$};
\end{scope}
\end{tikzpicture}
\caption{\label{fig:BSF1_scalar} 
The contribution of the scalar couplings to the radiative part of bound-state formation via emission of one scalar mediator. For fermionic $X_1,X_2$, only the diagram to the right exists.}
\end{figure}

The leading order diagrams in $\lambda_{1\varphi}$, $\lambda_{2\varphi}$ and $\rho_\varphi$ that contribute to the radiative part of the transition amplitude due to the scalar couplings in \cref{eq:L_Scalars_Real,eq:L_Scalars_Complex,eq:L_Fermions}  are shown in \cref{fig:BSF1_scalar}. We find
\begin{align}
i {\cal A}_T^\SC 
&= (-i\lambda_{1\varphi}) \frac{i}{(\eta_2 P_\varphi - q+p)^2 - m_\varphi^2} (-i g_2 m_2) 
\nonumber \\
&+ (-i\lambda_{2\varphi}) \frac{i}{(\eta_1 P_\varphi + q-p)^2 - m_\varphi^2} (-i g_1 m_1)
\nonumber \\
&+ 
(-i g_1 m_1) 
\frac{i}{(\eta_1 P_\varphi + q-p)^2 - m_\varphi^2} \ (-i\rho_\varphi) \ 
\frac{i}{(\eta_2 P_\varphi - q+p)^2 - m_\varphi^2}
(-i g_2 m_2) \,.
\end{align}  
Then according to the discussion in \cref{sec:BSF_general_RadiativeTransAmpl} on the scaling of the momenta,
\begin{equation}
{\cal A}_T^\SC \simeq 
\frac
{(\lambda_{1\varphi} g_2 \eta_2 + \lambda_{2\varphi} g_1 \eta_1) M}
{({\bf q}-{\bf p})^2 + m_\varphi^2}
- \frac{g_1g_2 M\mu \,  \rho_\varphi}{[({\bf q}-{\bf p})^2 + m_\varphi^2]^2} \,.
\label{eq:BSF1_ASC_scalarX}
\end{equation}
The contribution from the diagrams of \cref{fig:BSF1_trilinear} is~\cite{Petraki:2015hla,Petraki:2016cnz}
\begin{equation}
{\cal A}_T^{\TC} \simeq -2 M \mu \left[
g_1 (2\pi)^3 \delta^3({\bf q-p} - \eta_2 {\bf P_\varphi}) + 
g_2 (2\pi)^3 \delta^3({\bf q-p} + \eta_1 {\bf P_\varphi})
\right] \,.
\label{eq:BSF1_ATC_scalarX}
\end{equation}
Inserting \cref{eq:BSF1_ASC_scalarX,eq:BSF1_ATC_scalarX} in \cref{eq:MBSF}, and neglecting the corrections arising from the relativistic normalisation of states, we find
%
\begin{align}
{\cal M}_{{\bf k} \to \{n \ell m\}}^\SC &\simeq 
\frac{M}{\mu} \left[
\frac
{\lambda_{1\varphi} g_2 \eta_2 + \lambda_{2\varphi} g_1 \eta_1)}
{\sqrt{16 \pi \alpha_{sc}}}
\, {\cal V}_{{\bf k},\{n\ell m\}} 
-
\left(\frac{\rho_\varphi}{\mu \alpha_{sc}^2/2}\right)
\sqrt{4\pi\alpha_{sc}}
\, {\cal R}_{{\bf k},\{n\ell m\}} 
\right] ,
\label{eq:BSF1_MSC_scalarX}
\\
{\cal M}^{\TC}_{{\bf k}\rightarrow \{nlm\}} &\simeq
-M \sqrt{2\mu}\left[
g_1 {\cal I }_{{\bf k},\{nlm\}}(\eta_2 {\bf P}_\varphi)
+g_2 {\cal I }_{{\bf k},\{nlm\}}(-\eta_1 {\bf P}_\varphi)\right] \,.
\label{eq:BSF1_MTC_scalarX}
\end{align}
\Cref{eq:BSF1_MTC_scalarX} of course agrees with \cref{eq:MBSF1_nondeg}

\subsubsection{Cross-section for capture into the ground state \label{sec:BSF_SC_1phi_ScalarX_sigma}}

The cross-section for capture into the ground state is
\begin{equation}
\vrel \, \frac{d\sigma_{\BSFone}}{d\Omega} \simeq \frac{|{\bf P_\varphi}|}{64\pi^2 M^2 \mu }
|{\cal M}_{{\bf k} \to \{100\}}|^2 \,.
\label{eq:BSF1_dsigma}
\end{equation}
The momentum of the emitted scalar is found from the conservation of energy [cf.~\cref{eq:omega}]
\begin{equation}
\sqrt{ {\bf P}_\varphi^2 + m_\varphi^2 } \simeq
\frac{\mu}{2} (\alpha^2 + \vrel^2) \,,
\label{eq:BSF1_Pphi}
\end{equation}
where we adopted the Coulomb value for the binding energy since $\xi \gg 1$.

Taking into account the contributions of \cref{eq:BSF1_MTC_scalarX,eq:BSF1_MSC_scalarX} to the amplitude, and the overlap integrals of  \cref{eq:I_GroundState_re,eq:V_GroundState_re,eq:R_GroundState_re} for capture into the ground state, we obtain
\begin{align}
\sigma_{\BSFone}^{\rm tot} \vrel 
&\simeq
\frac{\alpha^2_{sc}}{\mu^2}
\, s_{ps}^{\BSFone} 
\left(\frac{\zeta^2}{1+\zeta^2}\right) 
e^{-4\zeta {\rm arccot} \, \zeta} 
\nonumber \\
&\times
\left\{
\left[ 
\frac{\lambda_{1\varphi} \, g_2 \eta_2 + \lambda_{2\varphi} \, g_1 \eta_1}{\sqrt{64 \pi^2 \alpha_{sc}}}
-
\sqrt{\alpha_{sc}}
\left(\frac{\rho_\varphi}{\mu \alpha^2_{sc}/2}\right)
\left(\frac{\zeta^2}{1+\zeta^2}\right)
\right]^2 S_0(\zeta,\xi)
\right.   
\nonumber \\  
&\left. 
+ \frac{2^6\pi}{3}
\frac{(g_1\eta_2-g_2\eta_1)^2}{16\pi \alpha_{sc}}
\left(\frac{\zeta^2}{1+\zeta^2} \right) \frac{S_1(\zeta,\xi)}{(1+\zeta^2)}
\right\} 
\,,
\label{eq:BSF1_sigma_scalarX}
\end{align}
where $s_{ps}^{\BSFone}$ is the phase-space suppression factor for BSF1,
\begin{equation}
s_{ps}^{\BSFone} \equiv 
\left(1- \left[\frac{2m_\varphi}{\mu (\alpha^2+\vrel^2)}\right]^2\right)^{1/2} \,.
\label{eq:BSF1_PhaseSpaceSuppr}
\end{equation}

As can be seen from \cref{eq:BSF1_sigma_scalarX}, the $\rho_\varphi$ contribution to the BSF1 cross-section scales as $\sigma_{\BSFone} \vrel~\propto~(1/\alpha)(\rho_{\varphi}/\mu)^2$. This is clearly of lower order in $\alpha$ than the cross-sections of \cref{sec:BSF_general_1phi_TC}, however it is suppressed by the square of the ratio of the mediator scale to the DM scale. Assuming that $\rho_\varphi \sim m_\varphi$, the kinematic threshold $m_\varphi \lesssim \mu \alpha^2/2$ implies that this contribution scales at best as $\sigma_{\BSFone} \vrel~\propto~\alpha^3$. While this is of higher order in $\alpha$ than \cref{eq:sigmaBSF1_nondeg}, it is still of lower order than \cref{eq:sigmaBSF1_deg,eq:sigmaBSF1_degF}. It is of course important to keep in mind that $\rho_\varphi$ may differ significantly from $m_\varphi$ and/or the binding energy.\footnote{\label{foot:rhophi}
We note that the third diagram of \cref{fig:BSF1_scalar}, from where the $\rho_\varphi$ contribution to BSF1 arises, resembles the diagram that appears in the radiative capture via one gluon emission in non-Abelian theories due to the trilinear gluon coupling~\cite{Asadi:2016ybp}. That coupling is momentum dependent, with the relevant momentum scale in the capture process being $\kappa$. We may recover the scaling of the non-Abelian BSF cross-section on $\alpha$~\cite{Harz:2018csl} by mapping $\rho_\varphi \to g\, \kappa~\propto~\alpha^{3/2}$.
}

We shall now simplify and adapt the above expression -- which has been derived for a pair of distinguishable scalars -- to various cases.

\subsubsection*{Different scalars species}

Let us consider for simplicity the limiting case where the couplings of the two particles are equal, $\lambda_{1\varphi} = \lambda_{2\varphi} = \lambda_{X\varphi}$ and $g_1 = g_2 = g$, while their masses are very different, $\eta_1 \gg \eta_2$. Then, \cref{eq:BSF1_sigma_scalarX} simplifies to
\begin{align}
&\sigma_{\BSFone}^{\rm tot} \vrel 
\simeq
\frac{\alpha^2_{sc}}{\mu^2}
s_{ps}^{\BSFone}
\left(\frac{\zeta^2}{1+\zeta^2}\right) 
e^{-4\zeta {\rm arccot} \, \zeta}
\nonumber \\
&\times
\left\{
\left[ 
\frac{\lambda_{X\varphi}}{\sqrt{4 \pi}}
- 
\sqrt{\alpha_{sc}}
\left(\frac{\rho_\varphi}{\mu \alpha^2_{sc}/2}\right)
\left(\frac{\zeta^2}{1+\zeta^2}\right)
\right]^2 
S_0(\zeta,\xi)
+ \frac{2^6\pi}{3}\left(\frac{\zeta^2}{1+\zeta^2} \right) \frac{S_1(\zeta,\xi)}{(1+\zeta^2)}
\right\} .
\label{eq:BSF1_sigma_scalarX_differentspecies}
\end{align}
In the regime where BSF is important and the Coulomb approximation is valid, i.e.~for $m_\varphi \lesssim \mu \vrel \lesssim \mu \alpha_{sc}$, we recall that $S_1 \simeq (1+\zeta^2) S_0$, thus all contributions exhibit the same velocity scaling, $\sigma \vrel~\propto~1/\vrel$. 
For perturbative $\lambda_{X\varphi}$ and for $\rho_\varphi \sim m_\varphi \lesssim \mu \alpha_{sc}^2/2$, the $s$-wave contributions are subdominant, 
and we recover the cross-section arising from the diagrams of \cref{fig:BSF1_trilinear} [cf.~\cref{eq:sigmaBSF1_nondeg}].

However, outside the Coulomb regime, i.e.~for $\mu \vrel \lesssim m_\varphi \lesssim \mu \alpha_{sc}$, the $p$-wave term becomes velocity suppressed (cf.~discussion in the end of \cref{sec:BSF_SC_OverlapIntegrals}). This implies that at sufficiently low velocities, the $\lambda_{X\varphi}$ and $\rho_\varphi$ contributions dominate.

\subsubsection*{Particle-antiparticle pair}
For a particle-antiparticle pair,	
$\lambda_{1\varphi} = \lambda_{2\varphi} \equiv \lambda_{X\varphi}$, 
$g_1 = g_2 \equiv g$ and
$m_1 = m_2 \equiv m_X$ (or $\eta_1 = \eta_2$). 
In this case, the $p$-wave term in \cref{eq:BSF1_sigma_scalarX} vanishes. The contribution from the $\lambda_{X\varphi}$ and $\rho_\varphi$ couplings becomes
\begin{align}
\sigma_{\BSFone}^{\SC} \vrel \simeq
\frac{\alpha_{sc}^2}{\mu^2}
\left[ \frac{\lambda_{X\varphi}}{\sqrt{4\pi}} 
- \sqrt{\alpha_{sc}}
\left(\frac{\rho_\varphi}{\mu \alpha_{sc}^2/2}\right)
\left(\frac{\zeta^2}{1+\zeta^2}\right)
\right]^2
s_{ps}^{\BSFone}
\ S_0(\zeta,\xi)
\left(\frac{\zeta^2}{1+\zeta^2}\right) 
e^{-4\zeta {\rm arccot} \, \zeta} .
\label{eq:BSF1_sigma_scalarX_XXdagger}
\end{align}
However, as discussed in \cref{sec:BSF_general_1phi_TC}, the trilinear $\varphi XX^\dagger$ coupling gives rise also to $s$- and $d$-wave contributions that are suppressed by higher orders in $\alpha_{sc}$~\cite{Petraki:2016cnz}. Comparing \cref{eq:BSF1_sigma_scalarX_XXdagger} with \cref{eq:sigmaBSF1_deg} in the Coulomb regime, we find the following.
\begin{itemize}

\item	
The $\lambda_{X\varphi}$ contribution to \cref{eq:BSF1_sigma_scalarX_XXdagger} dominates over \cref{eq:sigmaBSF1_deg} if
\begin{equation}
\lambda_{X\varphi} \gtrsim 18 \alpha_{sc} \,.
\label{eq:Condition_lambdaX}
\end{equation} 
Note that if the mediator $\varphi$ is the radial component of a complex scalar $\Phi$ that obtains a vacuum expectation value $v_\varphi$ and breaks a local symmetry, i.e.~$\Phi = (v_\varphi + \varphi)e^{i a_\varphi}/\sqrt{2}$, then the trilinear DM-DM-mediator coupling $g$ arises from the quartic coupling of this scalar to DM after spontaneous symmetry breaking, i.e.~
\begin{equation}
\delta {\cal L} = -\lambda_{X\varphi} |\Phi|^2 |X|^2 
\supset  
- \lambda_{X\varphi} \, v_\varphi \, \varphi |X|^2
- \frac{\lambda_{X\varphi}}{2} \, \varphi^2 |X|^2 \,.
\label{eq:L_SSB}
\end{equation} 
In this case, $\lambda_{X\varphi}$ and $g$ are related via $\lambda_{X\varphi} v_\varphi = g m_X$. Note that $m_X$ receives a contribution from $v_\varphi$, but remains an independent parameter. The above implies $\alpha_{sc} = \lambda_{X\varphi}^2 v_\varphi^2 / (16\pi m_X^2)$, and the condition \eqref{eq:Condition_lambdaX} becomes $\lambda_{X\varphi} \lesssim (16\pi/18) (m_X/v_\varphi)^2$, which encompasses the entire regime where $\lambda_{X\varphi}$ is perturbative if $v_\varphi \sim m_\varphi \ll m_X$ (or more generally, if $v_\varphi \lesssim m_X$). 
In this case, the $\lambda_{X\varphi}$ contribution to BSF1 dominates. 

\item

The $\rho_\varphi$ contribution to \cref{eq:BSF1_sigma_scalarX_XXdagger} dominates over \cref{eq:sigmaBSF1_deg} if 
\begin{equation}
\frac{\rho_\varphi }{ \mu \alpha_{sc}^2/2 } \gtrsim 3 \sqrt{\alpha_{sc}} \,,
\label{eq:Condition_rhophi_scalarX}
\end{equation}
which encompasses significant parameter space.

\end{itemize}
Note that outside the Coulomb regime, the  $\lambda_{X\varphi}$ and  $\rho_\varphi$ contributions dominate over \cref{eq:sigmaBSF1_deg} in a broader parameter ranges than those designated by the conditions \eqref{eq:Condition_lambdaX} and \eqref{eq:Condition_rhophi_scalarX}, since the $d$-wave component of  \cref{eq:sigmaBSF1_deg} becomes suppressed at low $\vrel$ while the $s$-wave terms saturate to constant values.

\subsubsection*{Identical scalars}

For a pair of identical scalars, the total wavefunction has to be symmetric in the interchange of the two particles. Thus the scattering state wavefunction is
\begin{equation}
\left[ \phi_{\bf k} (\bf r) + \phi_{-\bf k} (\bf r) \right] /\sqrt{2} \, ,
\label{eq:IdenticalScalars_SymWF}
\end{equation}
This implies that the contribution of the even-$\ell_S$ modes participating in a process is doubled with respect to the case of distinguishable scalars, while the contribution of the odd-$\ell_S$ modes vanishes. Similarly, there are only $\ell=$~even bound states of two identical bosons.
Since \cref{eq:BSF1_sigma_scalarX_XXdagger} includes only $s$-wave terms, the cross-section for a pair of identical scalars is twice as large as that given by \cref{eq:BSF1_sigma_scalarX_XXdagger}.

\subsection{Capture via emission of one scalar mediator $\varphi$, fermionic $X_1$ and $X_2$}
\label{sec:BSF_SC_1phi_FermionX}

\subsubsection{Amplitude}

For fermions, there are no renormalisable $\lambda_{j\varphi}$ couplings and only the third diagram in \cref{fig:BSF1_scalar} contributes. 
The amplitude is related to that for scalars as follows. The Dirac propagators $S_j^D$ can be expressed in terms of the scalar propagators $S_j (p)$ and the spinors $u, \bar{u}$
\begin{equation}
S_j^D (p) = \frac{i(\slashed{p}+m_j)}{p^2-m_j^2} 
= S_j(p) (\slashed{p}+m_j)
= S_j(p) \sum_{r} u_j^r(p) \bar{u}_j^r(p)  \,,
\label{eq:DiracProp}
\end{equation}
where $r$ denotes the spin and $j=1,2$ refers to the particle species. In order to compute the fermionic BSF diagrams, we insert a factor 
$\sum_{r_i} u_j^{r_i}(p_i) \bar{u}_j^{r_i} (p_i)$ 
for each propagator and contract 
$\bar{u}_j^{r_{i+1}} (p_{i+1}) u_j^{r_i} (p_i)$ across each vertex $i$. Since all the fermion-fermion-scalar vertices in the BSF diagrams are either soft or ultrasoft, we can use 
the identity 
\begin{equation}
\bar{u}_j^{s_{i+1}} (p_{i+1}) u^{s_i} (p_i) \simeq 
\bar{u}_j^{s_{i+1}} (p_i) u_j^{s_i} (p_i) 
=+2m_j \delta^{s_i s_{i+1}} \,,
\label{eq:ubaru_sec}
\end{equation}
as also in \cref{App:potential}. The spin Kronecker deltas in \cref{eq:ubaru_sec}, upon summation over the internal spin indices,  ensure that the spin of each particle is conserved across the entire diagram, including both the ladders and the vertices in the radiative parts of the diagrams. With this, we find
\begin{align}
i {\cal A}_T^\SC 
&= (-i g_1) 
\frac{i\bar{u}_1^{r_1'}(\eta_1K+q) u_1^{r_1}(\eta_1P+p)}
{(\eta_1 P_\varphi + q-p)^2 - m_\varphi^2} \ (-i\rho_\varphi) 
\ \frac{i\bar{u}_2^{r_2'}(\eta_1K+q) u_2^{r_2}(\eta_1P+p)}
{(\eta_2 P_\varphi - q+p)^2 - m_\varphi^2}
(-i g_2) 
\nonumber \\
&= \frac{-i 4M\mu g_1 g_2 \rho_\varphi}{[({\bf q-p})^2 + m_\varphi^2]^2} 
\ \delta^{r_1r_1'} \, \delta^{r_2r_2'} \,,
\label{eq:BSF1_ASC_fermionX}
\end{align} 
where $r_1, r_1'$ and $r_2, r_2'$ are the spins of the incoming and outgoing $X_1$ and $X_2$ particles. 
Comparing \cref{eq:BSF1_ASC_fermionX} to \eqref{eq:BSF1_ASC_scalarX}, we see that the $\rho_\varphi$ contribution to ${\cal A}_T$ is larger by a factor 4 for a fermionic pair than for a scalar pair. However, taking into account that for fermions $\alpha_f =g_1 g_2 / (4\pi)$, the full amplitude for fermions looks the same as the $\rho_\varphi$ contribution to \cref{eq:BSF1_MSC_scalarX}, up to the spin conservation factors,
\begin{equation}
{\cal M}_{{\bf k} \to \{n \ell m\}}^\SC \simeq 
- \frac{M}{\mu}
\left(\frac{\rho_\varphi}{\mu \alpha_f^2/2}\right)
\sqrt{4\pi\alpha_f}
\, {\cal R}_{{\bf k},\{n\ell m\}}  
\ \delta^{r_1r_1'} \, \delta^{r_2r_2'} \,,
\label{eq:BSF1_MSC_fermionX}
\end{equation} 
where we used \cref{eq:MBSF,eq:Overlap_R_def}. Similarly, the contribution to the amplitude from the diagrams of \cref{fig:BSF1_trilinear} is
\begin{equation}
{\cal M}^{\TC}_{{\bf k}\rightarrow \{nlm\}} \simeq
-2M \sqrt{2\mu}  \left[
g_1 {\cal I }_{{\bf k},\{ n\ell m\} }(\eta_2 {\bf P}_\varphi) +
g_2 {\cal I }_{{\bf k},\{ n\ell m\} }(-\eta_1 {\bf P}_\varphi) 
\right]
\, \delta^{r_1r_1'} \, \delta^{r_2r_2'} \,,
\label{eq:BSF1_MTC_fermionX}
\end{equation} 
where the extra factor $2$ with respect to \cref{eq:BSF1_MTC_scalarX} arises from the spinor contraction along the leg that contains the radiative vertex.

\subsubsection{Cross-section for capture into the ground state  \label{sec:BSF_SC_1phi_FermionX_sigma}}

Upon squaring the amplitudes \eqref{eq:BSF1_MSC_fermionX} and \eqref{eq:BSF1_MTC_fermionX}, summing over the final-state spins and averaging over the spin of the initial particles, the spin factors simply yield 1. 
Using \cref{eq:BSF1_dsigma,eq:BSF1_Pphi} and the overlap integrals \eqref{eq:R_GroundState_re} and \eqref{eq:I_GroundState_re}, we find the spin-averaged BSF1 cross-section  to be
\begin{align}
\sigma_{\BSFone}^{\rm tot} \vrel 
&\simeq
\frac{\alpha^2_{f}}{\mu^2}
\, s_{ps}^{\BSFone}
\left(\frac{\zeta^2}{1+\zeta^2}\right)^2 
e^{-4\zeta {\rm arccot} \, \zeta}
\nonumber \\
&\times \left\{
\alpha_{f}
\left(\frac{\rho_\varphi}{\mu \alpha_f^2/2}\right)^2
\left(\frac{\zeta^2}{1+\zeta^2}\right)
S_0(\zeta,\xi)
+ \frac{2^6 \pi}{3}
\frac{(g_1\eta_2-g_2\eta_1)^2}{4\pi \alpha_{f}}
\frac{S_1(\zeta,\xi)}{(1+\zeta^2)}
\right\} .
\label{eq:BSF1_sigma_fermionX}
\end{align}
As in the previous section, we shall now consider some specific cases.

\subsubsection*{Different fermion species}

We consider again the limiting case $g_1=g_2 \equiv g$ and $\eta_1 \gg \eta_2$. \Cref{eq:BSF1_sigma_fermionX} simplifies to
\begin{align}
\sigma_{\BSFone}^{\rm tot} \vrel 
&\simeq
\frac{\alpha^2_{f}}{\mu^2}
\, s_{ps}^{\BSFone}
\left(\frac{\zeta^2}{1+\zeta^2}\right)^2 
e^{-4\zeta {\rm arccot} \, \zeta}
\nonumber \\ 
&
\times \left\{
\alpha_{f}
\left(\frac{\rho_\varphi}{\mu \alpha_f^2/2}\right)^2
\left(\frac{\zeta^2}{1+\zeta^2}\right)
S_0(\zeta,\xi)
+ \frac{2^6 \pi}{3}
\frac{S_1(\zeta,\xi)}{(1+\zeta^2)}
\right\} .
\label{eq:BSF1_sigma_fermionX_differentspecies}
\end{align}
As in the case of scalar $X_{1,2}$, in the regime $m_\varphi < \mu \vrel < \mu \alpha$ where the Coulomb approximation holds, $S_1 = (1+\zeta^2)S_0$, and both terms in \cref{eq:BSF1_sigma_fermionX_differentspecies} scale as $\sigma \vrel~\propto~1/\vrel$. For $\rho_\varphi \sim m_\varphi \lesssim \mu \alpha_f^2/2$, the $s$-wave term is subdominant for perturbative $\alpha_{f}$.  However, at lower velocities $\vrel < m_\varphi / \mu$, the $p$-wave term dwindles and the $\rho_\varphi$ contribution dominates.

\subsubsection*{Particle-antiparticle pair}
For a particle-antiparticle pair, the $p$-wave contribution in \cref{eq:BSF1_sigma_fermionX} vanishes, and the BSF cross-section becomes
\begin{equation}
\sigma_{\BSFone}^\SC \vrel = 
\frac{\alpha_f^3}{\mu^2}
\left(\frac{\rho_\varphi}{\mu \alpha_f^2/2}\right)^2 
\ s_{ps}^{\BSFone} 
\times S_0(\zeta,\xi)
\left(\frac{\zeta^2}{1+\zeta^2}\right)^3 
e^{-4\zeta \, {\rm arccot} \, \zeta} \,.
\label{eq:BSF1_sigma_fermionX_XXbar}
\end{equation}
We recall that there are also $s$- and $d$-wave contributions of higher order in $\alpha_f$ from the trilinear $\varphi X \bar{X}$ coupling alone. Comparing \cref{eq:BSF1_sigma_fermionX_XXbar} with \cref{eq:sigmaBSF1_degF}, we find that the $\rho_\varphi$ term is more significant if 
\begin{equation}
\frac{\rho_\varphi }{ \mu \alpha_f^2/2 } \gtrsim 2.4 \sqrt{\alpha_f} \,.
\label{eq:Condition_rhophi_fermionX}
\end{equation}

\subsubsection*{Identical fermions}
For a pair of identical fermions, the total wavefunction has to be antisymmetric in the interchange of the two particles. This implies that the spatial wavefunction depends on their total spin. A pair of identical spin-1/2 particles may be either in the antisymmetric spin-0 state, or in the symmetric spin-1 state. Their spatial wavefunction should then be symmetric or antisymmetric, respectively,
\begin{align}
\text{fermions with total spin 0:}& \qquad
\left[ \phi_{\bf k} (\bf r) + \phi_{-\bf k} (\bf r) \right] /\sqrt{2} \, ,
\label{eq:IdenticalFermions_SymWF}
\\
\text{fermions with total spin 1:}& \qquad
\left[ \phi_{\bf k} (\bf r) - \phi_{-\bf k} (\bf r) \right] /\sqrt{2} \, .
\label{eq:IdenticalFermions_AntisymWF}
\end{align}
As for identical bosons, the wavefunction \eqref{eq:IdenticalFermions_SymWF} implies that the contribution of the even-$\ell_S$ modes participating in a process is doubled with respect to the case of distinguishable particles, while the contribution of the odd-$\ell_S$ modes vanishes. The opposite holds for the wavefunction \eqref{eq:IdenticalFermions_AntisymWF}. 
Similarly, there are only $\ell=$~even bound states of two identical total-spin-0 fermions, and only $\ell=$~odd bound states of two identical total-spin-one fermions. 
Since \cref{eq:BSF1_sigma_fermionX_XXbar} is an $s$-wave process, the spin-averaged cross-section for capture of two identical fermions is half of that given in 
\cref{eq:BSF1_sigma_fermionX_XXbar}, with the entire contribution arising from the spin-0 state.

\subsection{Capture via emission of two scalar mediators \label{sec:BSF_SC_2phi}}

The scalar couplings appearing in \cref{eq:L_Scalars_Real,eq:L_Scalars_Complex,eq:L_Fermions} raise the possibility that the radiative part of the capture into bound states may carry a lower suppression in $\alpha$ if two mediators are emitted. On the other hand, the emission of two mediators, which share the available energy $\omega \simeq \mu (\alpha^2 + \vrel^2)/2$, implies that the BSF2 cross-section picks up additional suppression in powers of $\alpha$ (for $\alpha > \vrel$) due the phase space of the second mediator. To determine which diagrams may contribute significantly, we first work out the phase-space integration.\footnote{
In a previous version of this paper, it was stated that the contributions to BSF2 from the trilinear DM-DM-mediator couplings $g_j$ alone do not suffer from the cancellations of the $g_j$ contributions to BSF1 shown in \cref{fig:BSF1_trilinear}. However, the resummation of the one-scalar-exchange diagrams between the two emission vertices in the corresponding BSF2 diagrams reveals that the same cancellations are present. Thus, these contributions to BSF2 are not of interest, and we do not consider them here. We thank Joan Soto for pointing this out.}

\subsubsection{Phase-space integration \label{sec:BSF_SC_2phi_PhaseSpace}}

The cross-section for capture via emission of two mediators with momenta ${\bf P}_a$ and ${\bf P}_b$ is 
\begin{equation}
d\sigma_{\BSFtwo} \! = \! \frac{1}{2E_12E_2 \vrel}
\frac{1}{2} \int \!
\frac{d^3P}{(2\pi)^3 2P^0}
\frac{d^3P_a}{(2\pi)^3 2P_a^0}
\frac{d^3P_b}{(2\pi)^3 2P^0_b} 
\abs{\mathcal{M}_{{\bf k}\to \{n\ell m\}}}^2
(2\pi)^4\delta^4(K-P-P_a-P_b) ,
\label{eq:dsigma}
\end{equation}
where the factor $(1/2)$ is due to the two identical particles in the final state. We work in the CM frame, ${\bf K=0}$, and use the three-momentum delta function to integrate over the momentum of the bound state ${\bf P}=-({\bf P}_a+{\bf P}_b)$. The energy delta function yields the condition
\begin{equation}
\sqrt{{\bf P}_{a}^2 + m_\varphi^2} + \sqrt{{\bf P}_{b}^2 + m_\varphi^2} + \frac{\bf P^2}{2M}
\simeq \epsilon_{n\ell}+\epsilon_{\bf k} .
\label{pss}
\end{equation}
For convenience, we define the dimensioneless parameters
\begin{equation}
\label{xaxb}
x_a = \frac{\abs{{\bf P}_{a}}}{\epsilon_{n\ell}+\epsilon_{\bf k}}, \qquad  
x_b = \frac{\abs{{\bf P}_{b}}}{\epsilon_{n\ell}+\epsilon_{\bf k}}
\end{equation}
and 
\begin{equation}
\delta\equiv\frac{\epsilon_{n\ell}+\epsilon_{\bf k}}{2M}, \qquad d\equiv\frac{m_\varphi}{\epsilon_{n\ell}+\epsilon_{\bf k}} \,.
\label{eq:deltad}
\end{equation}
Then, combining energy and momentum conservation, we obtain the phase space condition for the momenta of the emitted scalars,
\begin{equation}
\sqrt{x_a^2+d^2}+\sqrt{x_b^2+d^2}  
+ (x_a^2 + x_b^2 + 2x_a x_b \cos\theta_{ab})\delta = 1 \,,
\label{eq:PhaseSpaceCondition}
\end{equation}
where $\theta_{ab}$ is the angle between ${\bf P}_a$ and ${\bf P}_b$. Because $\delta \ll 1$, the phase-space encompassed by \cref{eq:PhaseSpaceCondition} extends on the $x_a-x_b$ plane essentially along the line 
\begin{equation} 
\sqrt{x_a^2+d^2} + \sqrt{x_b^2+d^2} = 1 \,, 
\label{eq:PhaseSpaceCondition_Real}
\end{equation} 
with $0\leqslant x_a \leqslant \sqrt{1-2d}$, and a small width  along the $x_b$ direction, due to $\cos \theta_{ab}$ ranging in $[-1,1]$. 
Note that for the capture process to be kinematically allowed, $2d< 1$. 
The $x_b$ width can be estimated by differentiating \cref{eq:PhaseSpaceCondition} with respect to $x_b$ and $\cos \theta_{ab}$. We find
\begin{equation}
x_b^{\max} - x_b^{\min} \simeq 4 x_a \left(1-\sqrt{x_a^2+d^2}\right) \delta \,.
\label{eq:xbinterval}
\end{equation}

For the diagrams of interest, the amplitude is independent of ${\bf P}_{a,b}$ (cf.~\cref{sec:BSF_SC_2phi_AmplitudeCrossSection}). Thus, putting everything together, \cref{eq:dsigma} yields
\begin{equation}
\sigma_{\BSFtwo} \vrel \simeq  
\frac{(\epsilon_{n\ell}+\epsilon_{\bf k})^3}{2^8 3 \pi^3 M^2 \mu}
\ \abs{\mathcal{M}_{{\bf k}\rightarrow \{n \ell m\}}}^2
\ s_{ps}^{\BSFtwo} (d) \,,
\label{eq:sigmaBSF2}
\end{equation}
where $s_{ps}^{\BSFtwo}$ is the phase-space suppression factor due to the kinematic threshold for BSF2,
\begin{equation}
s_{ps}^{\BSFtwo} (d) \equiv 
6\int_{0}^{\sqrt{1-2d}}   d x \ x^2 
\ \sqrt{\frac{1+x^2-2\sqrt{x^2+d^2}}{x^2+d^2}} \,,
\label{eq:BSF2_PhaseSpaceSuppr}
\end{equation}
with $s_{ps}^{\BSFtwo} = 1$ for $d=0$. Comparing \cref{eq:sigmaBSF2} with \cref{eq:BSF1_dsigma}, we observe that the BSF2 cross-section is proportional to two extra powers of the available energy to be dissipated, $\omega \simeq \epsilon_{n\ell}+\epsilon_{\bf k}$, with respect to BSF1, as expected from the phase-space element for the second mediator $d^3P_b / (2P_b^0)~\propto~\omega^2$. This implies an extra suppression by $\alpha^4$ in the regime where BSF is important, $\alpha > \vrel$. We also note the suppression of BSF2 with respect to BSF1 by the numerical factor $1/(2^4 3\pi^2) \simeq 2 \times 10^{-3}$.

\subsubsection{Amplitude and cross-section \label{sec:BSF_SC_2phi_AmplitudeCrossSection}}

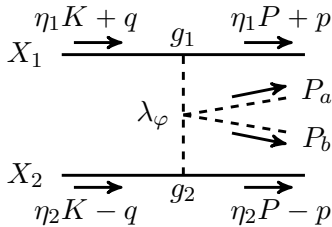
\begin{figure}[t]
\centering
\begin{tikzpicture}[line width=1.1pt, scale=1.6]
\begin{scope}[shift={(0,0)}]
\draw[fermionnoarrow] (-1,1) -- (1,1);
\draw[fermionnoarrow] (-1,0) -- (1,0);
\draw[scalarnoarrow]   (0,0) -- (0,1);
\draw[scalarnoarrow]   (0,0.5) -- (0.9,0.35);
\draw[scalarnoarrow]   (0,0.5) -- (0.9,0.65);
\node at (-1.3,1) {$X_1$};
\node at (-1.3,0) {$X_2$};
\draw[->] (-0.9, 1.1) -- (-0.5, 1.1);
\draw[->] ( 0.5, 1.1) -- ( 0.9, 1.1);
\draw[->] (-0.9,-0.1) -- (-0.5,-0.1);
\draw[->] ( 0.5,-0.1) -- ( 0.9,-0.1);
\node at (-0.8, 1.3) {$\eta_1 K + q$};
\node at ( 0.8, 1.3) {$\eta_1 P + p$};
\node at (-0.8,-0.3) {$\eta_2 K - q$};
\node at ( 0.8,-0.3) {$\eta_2 P - p$};
\draw[->] (0.4,0.65) -- (0.85,0.74);
\draw[->] (0.4,0.35) -- (0.85,0.26);
\node at (1.1,0.7) {$P_a$};
\node at (1.1,0.3) {$P_b$};
\node at ( 0, 1.14) {$g_1$};
\node at ( 0,-0.15) {$g_2$};
\node at (-0.25,0.5)  {$\lambda_\varphi$};
\end{scope}
\end{tikzpicture}
\caption{\label{fig:BSF2_MediatorQuartic} 
Capture via emission of two scalar mediators: the contribution of the mediator quartic self-coupling to the radiative amplitude.}
\end{figure}

In \cref{fig:BSF2_MediatorQuartic,fig:BSF2_MediatorTrilinear,fig:BSF2_lambdaj}, we show various contributions to BSF2 arising from the mediator self-couplings and the mediator-DM quartic coupling for scalar DM. The diagram of \cref{fig:BSF2_MediatorQuartic} yields the most important contribution. We shall first compute the cross-section arising from this diagram and then discuss why we neglect the other contributions.

The amplitude for the $\lambda_\varphi$ contribution can be obtained from the second term in \cref{eq:BSF1_MSC_scalarX} by replacing $\rho_\varphi$ with $\lambda_\varphi$. Taking into account the discussion in \cref{sec:BSF_SC_1phi_FermionX_sigma}, we find it to be the same for both bosons and fermions,
\begin{equation}
{\cal M}_{{\bf k}\to \{n\ell m\}}^\SC \simeq 
\lambda_{\varphi} \, \sqrt{\frac{16\pi}{\alpha^3}} \frac{M}{\mu^2} 
\ {\cal R}_{{\bf k},\{n \ell m \}} \,,
\label{eq:BSF2_scalar_M}
\end{equation}
where ${\cal R}_{{\bf k},\{n \ell m \}}$ is defined in \cref{eq:Overlap_R_def}. From \cref{eq:sigmaBSF2,eq:R_GroundState_re} we obtain the (spin-averaged) cross-section for capture into the ground state,
\begin{equation}
\sigma_{\BSFtwo}^\SC \vrel \simeq  f
\ \frac{\lambda_\varphi^2 \alpha^3}{48\pi^2 \mu^2}
\ s_{ps}^{\BSFtwo} 
\ S_0 \, \left(\frac{\zeta^2}{1+\zeta^2}\right) \, e^{-4\zeta\,{\rm arccot}\,\zeta} \,,
\label{eq:BSF2_SC_sigma}
\end{equation}
where $s_{ps}^{\BSFtwo}$ and  $S_0$ are given in eqs.~\eqref{eq:BSF2_PhaseSpaceSuppr} and \eqref{eq:Hulthen_S0_re} respectively. The factor $f$ follows from the discussion in \cref{sec:BSF_SC_1phi_ScalarX,sec:BSF_SC_1phi_FermionX}, with
\begin{equation}
f = \left\{
\begin{alignedat}{6}
&1,& \qquad &\text{distinguishable scalars,}& \\
&2,& \qquad &\text{identical scalars,}& \\
&1/4,& \qquad &\text{distinguishable fermions,}& \\
&1/2,& \qquad &\text{identical fermions.}&
\end{alignedat}
\right.	
\end{equation}
The cross-section \eqref{eq:BSF2_SC_sigma} becomes comparable to the $\rho_\varphi$ contribution to BSF1 computed in \cref{sec:BSF_SC_1phi_ScalarX_sigma,sec:BSF_SC_1phi_FermionX_sigma} for
\begin{equation}
\frac{\lambda_\varphi}{4\sqrt{3} \pi} \sim \frac{\rho_\varphi}{\mu \alpha^2/2} \,.
\label{eq:Condition_lambdaphiVSrhophi}
\end{equation}
Comparing with the BSF1 contributions from the trilinear coupling alone \eqref{eq:sigmaBSF1_deg} and \eqref{eq:sigmaBSF1_degF}, the $\lambda_\varphi$ contribution to BSF2 is more significant for
\begin{equation}
\lambda_\varphi \gtrsim 10^2 \sqrt{\alpha} \,.
\label{eq:Condition_lambdaphiVSalpha}
\end{equation}
The perturbativity of the couplings implies that the above condition may be meaningfully satisfied only for small coupling $\alpha \lesssim 10^{-3}$.

\subsubsection{Other subdominant contributions \label{sec:BSF_SC_2phi_Other}}

In \cref{fig:BSF2_MediatorTrilinear,fig:BSF2_lambdaj}, we display various diagrams arising from $\rho_\varphi$ and $\lambda_{j\varphi}$. Their contributions to BSF2 are subdominant compared to the contributions of the same couplings to BSF1 for reasonable choices of the parameters, as we discuss below. Since for phenomenological purposes, we are interested mostly in the total capture rate, we do not compute these contributions in detail.

Because part of the suppression arises from the phase space density of the second mediator (cf.~\cref{sec:BSF_SC_2phi_PhaseSpace}), we will consider directly the contributions of these diagrams to the cross-section rather than the amplitude, thus neglecting any cross-terms between different diagrams (except in the cases where there is a cancellation between them). It is straightforward to verify that the cross-terms do not alter our conclusions. We focus on particle-antiparticle or identical-particle pairs. Our comparisons refer to the regime $\alpha > \vrel$ where BSF is important, setting aside the $\zeta$-dependent factors that yield the same $\vrel$ dependence between BSF1 and BSF2 in this regime.

\begin{figure}
\begin{tikzpicture}[line width=1.1pt, scale=1.6]
\begin{scope}[shift={(-3.5,0)}]
\draw[fermionnoarrow] (-0.8,1) -- (0.8,1);
\draw[fermionnoarrow] (-0.8,0) -- (0.8,0);
\draw[scalarnoarrow]  (0,1.4) -- ( 0.3,1.9);
\draw[scalarnoarrow]  (0,1.4) -- (-0.3,1.9);
\draw[scalarnoarrow]  (0,1) -- (0,1.4);
\draw[->] ( 0.22, 1.55) -- ( 0.4, 1.85);
\draw[->] (-0.22, 1.55) -- (-0.4, 1.85);
%
\node at (-0.6, 1.7) {$P_a$};
\node at ( 0.6, 1.7) {$P_b$};
\node at (0, 0.8) {$g_1$};
\node at (0.2,1.3) {$\rho_\varphi$};
\end{scope}
\begin{scope}[shift={(-1.3,0)}]
\draw[fermionnoarrow] (-0.8,1) -- (0.8,1);
\draw[fermionnoarrow] (-0.8,0) -- (0.8,0);
\draw[scalarnoarrow]  (0,-0.4) -- ( 0.3,-0.9);
\draw[scalarnoarrow]  (0,-0.4) -- (-0.3,-0.9);
\draw[scalarnoarrow]  (0,0) -- (0,-0.4);
\draw[->] ( 0.22,-0.55) -- ( 0.4,-0.85);
\draw[->] (-0.22,-0.55) -- (-0.4,-0.85);
%
\node at (-0.6,-0.7) {$P_a$};
\node at ( 0.6,-0.7) {$P_b$};
\node at (0,  0.15) {$g_2$};
\node at (0.2,-0.3) {$\rho_\varphi$};
\end{scope}
\begin{scope}[shift={(1.3,0)}]
\draw[fermionnoarrow] (-0.8,1) -- (0.8,1);
\draw[fermionnoarrow] (-0.8,0) -- (0.8,0);
\draw[scalarnoarrow]   (0,0) -- (0,1);
\draw[scalarnoarrow]   (0,0.5) -- (0.4,0.5);
\draw[scalarnoarrow]   (0.4,0.5) -- (0.8,0.85);
\draw[scalarnoarrow]   (0.4,0.5) -- (0.8,0.15);
\draw[->] (0.5,0.72) -- (0.75,0.92);
\draw[->] (0.5,0.3) -- (0.75,0.08);
\node at (0.93,0.82) {$P_a$};
\node at (0.93,0.18) {$P_b$};
\node at ( 0, 1.14) {$g_1$};
\node at ( 0,-0.15) {$g_2$};
\node at (-0.25,0.5)  {$\rho_\varphi$};
\node at ( 0.65,0.5)  {$\rho_\varphi$};
\end{scope}
\begin{scope}[shift={(3.5,0)}]
\draw[fermionnoarrow] (-0.8,1) -- (0.8,1);
\draw[fermionnoarrow] (-0.8,0) -- (0.8,0);
\draw[scalarnoarrow]   (0,0) -- (0,1);
\draw[scalarnoarrow]   (0,0.7) -- (0.5,0.7);
\draw[scalarnoarrow]   (0,0.3) -- (0.5,0.3);
\draw[->] (0.2,0.62) -- (0.5,0.62);
\draw[->] (0.2,0.38) -- (0.5,0.38);
\node at (0.76,0.62) {$P_{a/b}$};
\node at (0.76,0.35) {$P_{b/a}$};
\node at ( 0, 1.14) {$g_1$};
\node at ( 0,-0.15) {$g_2$};
\node at (-0.25,0.7)  {$\rho_\varphi$};
\node at (-0.25,0.3)  {$\rho_\varphi$};
\end{scope}
\end{tikzpicture}
\caption{\label{fig:BSF2_MediatorTrilinear} 
Capture via emission of two scalar mediators: the contributions of the mediator trilinear self-coupling to the radiative amplitude.}
%
%
%
%
\begin{tikzpicture}[line width=1.1pt, scale=1.6]
\begin{scope}[shift={(-4,0)}]
\draw[fermionnoarrow] (-0.6,1) -- (0.6,1);
\draw[fermionnoarrow] (-0.6,0) -- (0.6,0);
\draw[scalarnoarrow]  (0,-0.6) -- (0,1.6);
\draw[->] ( 0.1, 1.1) -- ( 0.1, 1.55);
\draw[->] ( 0.1,-0.1) -- ( 0.1,-0.55);
\node at ( 0.35, 1.45) {$P_a$};
\node at ( 0.35,-0.45) {$P_b$};
\node at ( 0.2, 0.8 ) {$\lambda_{1\varphi}$};
\node at ( 0.2, 0.15) {$\lambda_{2\varphi}$};
\end{scope}
\begin{scope}[shift={(-2.6,0)}]
\draw[fermionnoarrow] (-0.6,1) -- (0.6,1);
\draw[fermionnoarrow] (-0.6,0) -- (0.6,0);
\draw[scalarnoarrow]  (0,-0.6) -- (0,1.6);
\draw[->] ( 0.1, 1.1) -- ( 0.1, 1.55);
\draw[->] ( 0.1,-0.1) -- ( 0.1,-0.55);
\node at ( 0.35, 1.45) {$P_b$};
\node at ( 0.35,-0.45) {$P_a$};
\node at ( 0.2, 0.8 ) {$\lambda_{1\varphi}$};
\node at ( 0.2, 0.15) {$\lambda_{2\varphi}$};
\end{scope}
\begin{scope}[shift={(-0.7,0)}]
\draw[fermionnoarrow] (-0.6,1) -- (0.6,1);
\draw[fermionnoarrow] (-0.6,0) -- (0.6,0);
\draw[scalarnoarrow]  ( 0,  0) -- (0,1.6);
\draw[scalarnoarrow]  ( 0,0.5) -- (0.5,0.5);
\draw[->] ( 0.1, 1.1) -- ( 0.1, 1.55);
\draw[->] ( 0.1, 0.4) -- ( 0.45,0.4);
\node at ( 0.4, 1.45) {$P_{a/b}$};
\node at ( 0.4, 0.15) {$P_{b/a}$};
%
\node at ( 0.2, 0.8 ) {$\lambda_{1\varphi}$};
\node at ( 0,-0.15) {$g_{2}$};
\node at (-0.2, 0.5) {$\rho_\varphi$};
\end{scope}
\begin{scope}[shift={( 0.7,0)}]
\draw[fermionnoarrow] (-0.6,1) -- (0.6,1);
\draw[fermionnoarrow] (-0.6,0) -- (0.6,0);
\draw[scalarnoarrow]  ( 0, -0.6) -- (0,1);
\draw[scalarnoarrow]  ( 0,0.5) -- (0.5,0.5);
\draw[->] ( 0.1,-0.1) -- ( 0.1,-0.55);
\draw[->] ( 0.1, 0.6) -- ( 0.45,0.6);
\node at ( 0.4,-0.45) {$P_{a/b}$};
\node at ( 0.4, 0.8 ) {$P_{b/a}$};
%
\node at ( 0.2,0.15) {$\lambda_{2\varphi}$};
\node at ( 0,  1.15) {$g_1$};
\node at (-0.2, 0.5) {$\rho_\varphi$};
\end{scope}
\begin{scope}[shift={(2.6,0)}]
\draw[fermionnoarrow] (-0.6,1) -- (0.6,1);
\draw[fermionnoarrow] (-0.6,0) -- (0.6,0);
\draw[scalarnoarrow]  (0,1) -- ( 0.3,1.6);
\draw[scalarnoarrow]  (0,1) -- (-0.3,1.6);
\draw[->] ( 0.23, 1.2) -- ( 0.4, 1.55);
\draw[->] (-0.23, 1.2) -- (-0.4, 1.55);
%
\node at (-0.6, 1.45) {$P_a$};
\node at ( 0.6, 1.45) {$P_b$};
\node at (0, 0.8) {$\lambda_{1\varphi}$};
\end{scope}
\begin{scope}[shift={(4,0)}]
\draw[fermionnoarrow] (-0.6,1) -- (0.6,1);
\draw[fermionnoarrow] (-0.6,0) -- (0.6,0);
\draw[scalarnoarrow]  (0,0) -- ( 0.3,-0.6);
\draw[scalarnoarrow]  (0,0) -- (-0.3,-0.6);
\draw[->] ( 0.23,-0.2) -- ( 0.4,-0.55);
\draw[->] (-0.23,-0.2) -- (-0.4,-0.55);
%
\node at (-0.6,-0.45) {$P_a$};
\node at ( 0.6,-0.45) {$P_b$};
\node at (0, 0.2) {$\lambda_{2\varphi}$};
\end{scope}
\end{tikzpicture}
\caption{\label{fig:BSF2_lambdaj} 
Capture via emission of two scalar mediators: the contribution of the quartic DM-mediator coupling to the radiative amplitude. These diagrams exist for scalar DM only.}
\end{figure}

\begin{itemize}
\item	
For particle-antiparticle or identical-particle pairs, the sum of the two diagrams on the left in \cref{fig:BSF2_MediatorTrilinear} suffers from the same cancellation as the diagrams of \cref{fig:BSF1_trilinear} (cf.~\cref{sec:BSF_general_1phi_TC}). This introduces a $\alpha^4$ suppression in the corresponding cross-section, while the phase-space density of the second emitted mediator implies an additional suppression by $(\mu \alpha^2/2)^2$ as discussed in \cref{sec:BSF_SC_2phi_PhaseSpace}. These suppressions are balanced by the propagator of the off-shell mediator, which introduces a factor $\sim (P_a^0+P_b^0)^{-4}~\propto~(\mu \alpha^2/2)^{-4}$. Thus, the contribution of these diagrams to the cross-section scales as $\sigma_{\BSFtwo} \vrel~\propto~(\rho_\varphi/\mu)^2$. This is suppressed by $\alpha$ with respect to the $\rho_\varphi$ contribution to BSF1, $\sigma_{\BSFone} \vrel~\propto~(1/\alpha)(\rho_\varphi/\mu)^2$ [see e.g.~\cref{eq:BSF1_sigma_fermionX_XXbar}], as well as by the numerical factor $\sim 2\times 10^{-3}$ due to the three-body phase space. 
	
\item
The third and forth diagrams in \cref{fig:BSF2_MediatorTrilinear} can be estimated starting from the $\rho_\varphi$ contribution to BSF1 [cf.~third diagram in \cref{fig:BSF1_scalar} and \cref{eq:BSF1_sigma_fermionX_XXbar}]. 
	
For the third diagram, the off-shell mediator yields a factor $\propto~(\mu\alpha^2/2)^{-4}$. Taking into account the phase-space suppression $\propto~(\mu \alpha^2/2)^2$, we find that the contribution to the BSF2 cross-section scales as $\sigma_{\BSFtwo} \vrel~\propto~(1/\alpha^5)(\rho_\varphi/\mu)^4$. Assuming that $\rho_\varphi \sim m_\varphi \lesssim \mu \alpha^2/4$ (cf.~\cref{foot:rhophi}), effectively this yields at best $\sigma_{\BSFtwo} \vrel~\propto~\alpha^3$, similarly to the BSF1 cross-section \eqref{eq:BSF1_sigma_fermionX_XXbar}. Still, the numerical suppression of BSF2 due to the three-body phase space ensures that the BSF2 contribution is subdominant to BSF1. However, if there is a large hierarchy between $m_\varphi$ and $\rho_\varphi$ with $\rho_\varphi \gg m_\varphi$, then the diagram under consideration may be significant.\footnote{
We note that the three left diagrams in \cref{fig:BSF2_MediatorTrilinear} exhibit a collinear divergence at $m_\varphi \to 0$ that would have to be treated appropriately.}

For the forth diagram, the off-shell mediator between the two emitted scalars yields a factor $\propto~(\mu\alpha)^{-4}$. It follows that it is subdominant to the third diagram in  \cref{fig:BSF2_MediatorTrilinear} and to the $\rho_\varphi$ contribution to BSF1.

\item 
	
For the two diagrams on the left in \cref{fig:BSF2_lambdaj}, we start from the first term of \cref{eq:BSF1_sigma_scalarX_XXdagger}, replace one factor of $\alpha$ with $\lambda_{X\varphi}^2$ and introduce the phase-space suppression $\propto~\alpha^4$. We obtain $\sigma_{\BSFtwo}\vrel~\propto~\alpha^5\lambda_{X\varphi}^4$.

\item 
For the two middle diagrams in \cref{fig:BSF2_lambdaj}, we start from the second term of \cref{eq:BSF1_sigma_scalarX_XXdagger}, replace one power of $\alpha$ with $\lambda_{X\varphi}^2$ and introduce the phase-space suppression $\propto~\alpha^4$. We obtain $\sigma_{\BSFtwo}\vrel~\propto~\alpha^2\lambda_{X\varphi}^2 (\rho_\varphi/\mu)^2$, which is subdominant to the first term of \cref{eq:BSF1_sigma_scalarX_XXdagger} for $\rho_\varphi \ll \mu$.
	
\item 
The two diagrams on the right in \cref{fig:BSF2_lambdaj} suffer from the cancellations of the BSF1 diagrams of \cref{fig:BSF1_trilinear}. Replacing one factor of $\alpha$ with $\lambda_{X\varphi}^2$ results in the suppression $\lambda_{X\varphi}^2\alpha^3$ for particle-antiparticle or identical-particle pairs. Introducing also the phase-space suppression yields $\sigma_{\BSFtwo} \vrel~\propto~\lambda_{X\varphi}^2\alpha^7$.

\end{itemize}

\clearpage
\section{Conclusion \label{Sec:Conc}}

Dark matter coupled to a light scalar mediator is motivated in the context of Higgs portal models, self-interacting DM, and more generally models beyond the SM with extended scalar sectors. Light force mediators generically imply the existence of DM bound states, whose formation has important implications for the DM phenomenology. While the cosmological and astrophysical formation of bound states is significant in models with vector mediators, it is less efficient in models with scalars mediators. Indeed, considering only a trilinear DM-DM-mediator coupling, the radiative capture of particle-antiparticle or identical-particle pairs into bound states with emission of a scalar mediator is subject to cancellations that suppress the cross-section by higher powers in this coupling.

While the trilinear coupling determines the long-range interaction between the DM particles, it is not necessarily the only contributor to the radiative part of the transition. Quite generically, in theories with scalar mediators the couplings in the scalar potential also contribute. In this work, we investigated the contribution of the mediator self-couplings, as well as the DM-mediator quartic coupling in the case of scalar DM, to the radiative capture into bound states. We considered capture both via one and two scalar mediator emission.  Our main results include \cref{eq:BSF1_sigma_scalarX_differentspecies,eq:BSF1_sigma_scalarX_XXdagger} for BSF1 by scalar particles, \cref{eq:BSF1_sigma_fermionX_differentspecies,eq:BSF1_sigma_fermionX_XXbar} for BSF1 by fermionic particles, and in \cref{eq:BSF2_SC_sigma} for BSF2.  We have found that the newly considered couplings can enhance or dominate the capture rate in sizeable parts of the parameter space, described in part by \cref{eq:Condition_lambdaX,eq:Condition_rhophi_scalarX,eq:Condition_rhophi_fermionX,eq:Condition_lambdaphiVSrhophi,eq:Condition_lambdaphiVSalpha}.

Importantly, the new contributions are $s$-wave and remain significant even at very low velocities, thereby enhancing the radiative signals that can be probed in the indirect searches. This can potentially strengthen the resulting constraints. While models with light mediators and $s$-wave annihilation are severely constrained by the CMB and other indirect probes~\cite{Bringmann:2016din,Cirelli:2016rnw,Kahlhoefer:2017umn}, models with $p$-wave annihilation -- such as fermionic DM coupled to a scalar mediator -- remain largely unconstrained. In this case, the formation and subsequent decay of unstable bound states offers a source of detectable indirect signals~\cite{An:2016kie}. (Bremsstrahlung of dark mediators has also been invoked to lift the $p$-wave suppression~\cite{Bell:2017irk}.)
Moreover, $s$-wave annihilation processes may lead to a period of reannihilation in the early universe, after DM kinetic decoupling~\cite{Binder:2017lkj}.

Besides the signals produced from the decay of unstable bound states, the radiation emitted during the capture process is another source of indirect signals even in the case of stable bound states~\cite{Pearce:2013ola}. This is particularly important for asymmetric DM~\cite{Petraki:2013wwa}, whose annihilation signals are suppressed due to the asymmetry (albeit can still be significant due to the Sommerfeld effect~\cite{Baldes:2017gzw}). Asymmetric DM, in turn, offers an excellent host of self-interacting DM, both because it allows for large couplings to light mediators, and it evades the indirect detection constraints, provided that the asymmetry is sufficiently large~\cite{Baldes:2017gzu}. The enhanced radiative signals expected due to the contributions computed here can improve the prospects of probing asymmetric and self-interacting DM.

Moreover, for asymmetric DM coupled to a light scalar, the enhanced rate of formation of two-particle bound states can result in more efficient cosmological formation of multiparticle bound states~\cite{Wise:2014jva}, which in the case of scalar DM may lead to solitosynthesis~\cite{Kusenko:1997hj,Postma:2001ea,Pearce:2012jp}.

\clearpage
\appendix
\section*{Appendices}

\section{Non-relativistic potential for scalars and fermions  \label{App:potential}}
\input{appendix/Potential.tex}

\section{Scattering state and bound state wavefunctions  \label{App:wf}}
\input{appendix/Wavefunctions.tex}

\section{Overlap integrals  \label{App:OverlapIntegrals}}
\input{appendix/OverlapIntegral.tex}

\section*{Acknowledgments}
We thank Alex Kusenko, Andreas Goudelis, Marieke Postma, Enrico Russo and Joan Soto for helpful discussions. This work was supported by the NWO Vidi grant ``Self-interacting asymmetric dark matter" and by the ANR ACHN 2015 grant (``TheIntricateDark" project).

\clearpage
\bibliography{Bibliography.bib}

\end{document}

%% file: TIKZ_prelim.tex
\usepackage{tikz}
\usetikzlibrary{arrows,shapes}
\usetikzlibrary{trees,patterns}
\usetikzlibrary{matrix,arrows} 				
\usetikzlibrary{positioning}				  
\usetikzlibrary{calc,through}				  
\usetikzlibrary{decorations.pathreplacing}  
\usepackage{pgffor}							

\usetikzlibrary{decorations.pathmorphing}	
\usetikzlibrary{decorations.markings}
\tikzset{
	>=stealth', 
    vector/.style={decorate, decoration={snake}, draw},
	provector/.style={decorate, decoration={snake,amplitude=2.5pt}, draw},
	antivector/.style={decorate, decoration={snake,amplitude=-2.5pt}, draw},
	bigvector/.style={decorate, decoration={snake,amplitude=4pt}, draw},
    fermion/.style={draw=black, postaction={decorate},
        decoration={markings,mark=at position .55 with {\arrow[draw=black]{>}}}},
    fermionbar/.style={draw=black, postaction={decorate},
        decoration={markings,mark=at position .55 with {\arrow[draw=black]{<}}}},
    fermionnoarrow/.style={draw=black},
    gluon/.style={decorate, draw=black,
        decoration={coil,amplitude=4pt, segment length=5pt}},
    scalar/.style={dashed,draw=black, postaction={decorate},
        decoration={markings,mark=at position .55 with {\arrow[draw=black]{>}}}},
    scalarbar/.style={dashed,draw=black, postaction={decorate},
        decoration={markings,mark=at position .55 with {\arrow[draw=black]{<}}}},
    scalarnoarrow/.style={dashed,draw=black},
    momentum/.style={draw=black, postaction={decorate},
        decoration={markings,mark=at position 1 with {\arrow[draw=black]{>}}}},
    antimomentum/.style={draw=black, postaction={decorate},
        decoration={markings,mark=at position 0.1 with {\arrow[draw=black]{<}}}}
}

\tikzstyle{block} = [draw, rectangle, 
    minimum height=3em, minimum width=6em]

%% file: appendix/Potential.tex
\begin{figure}[h]
\centering
\begin{tikzpicture}[line width=1.5 pt, scale=1.8]
\begin{scope}[shift={(-2,0)}]
\draw[fermion] (-1, 0)--(0, 0);
\draw[fermion] ( 0, 0)--(1, 0);
\draw[fermion] (-1,-1)--(0,-1);
\draw[fermion] ( 0,-1)--(1,-1);
\draw[scalarnoarrow] (0,0)--(0,-1);
\draw[->] (0.2,-0.3) -- (0.2,-0.7);
\node at (-1.7, 0) {$X_1$};
\node at (-1.7,-1) {$X_2$};
\node at (-1.25, 0) {$(s_1)$};
\node at (-1.25,-1) {$(s_2)$};
\node at ( 1.25, 0) {$(s_1')$};
\node at ( 1.25,-1) {$(s_2')$};
\node at (-0.6, 0.25){$\eta_1 P + p$};
\node at (-0.6,-1.25){$\eta_2 P - p$};
\node at ( 0.6, 0.25){$\eta_1 P + p'$};
\node at ( 0.6,-1.25){$\eta_2 P - p'$};
\node at (0.8,-0.5){$k=p-p'$};
\end{scope}
\begin{scope}[shift={(2,0)}]
\draw[fermion] (-1, 0)--(0, 0);
\draw[fermion] ( 0, 0)--(1, 0);
\draw[fermionbar] (-1,-1)--(0,-1);
\draw[fermionbar] ( 0,-1)--(1,-1);
\draw[scalarnoarrow] (0,0)--(0,-1);
\draw[->] (-0.7, 0.2) -- (-0.3, 0.2);
\draw[->] (-0.7,-1.2) -- (-0.3,-1.2);
\draw[->] ( 0.3, 0.2) -- ( 0.7, 0.2);
\draw[->] ( 0.3,-1.2) -- ( 0.7,-1.2);
\draw[->] (0.2,-0.3) -- (0.2,-0.7);
\node at (-1.25, 0) {$(s_1)$};
\node at (-1.25,-1) {$(s_2)$};
\node at ( 1.25, 0) {$(s_1')$};
\node at ( 1.25,-1) {$(s_2')$};
\node at (-0.6, 0.4){$\eta_1 P + p$};
\node at (-0.6,-1.4){$\eta_2 P - p$};
\node at ( 0.6, 0.4){$\eta_1 P + p'$};
\node at ( 0.6,-1.4){$\eta_2 P - p'$};
\node at (0.8,-0.5){$k=p-p'$};
\end{scope}
\end{tikzpicture}
\caption{One boson exchange diagrams that yield the leading order contribution to the non-relativistic potential between two different paticles $X_1$ and $X_2$ (\emph{left}), or  a particle-antiparticle pair (\emph{right}). The parentheses denote the spin of the incoming and outgoing particles, in the case of fermions.}    
\label{fig:1BED}
\end{figure}
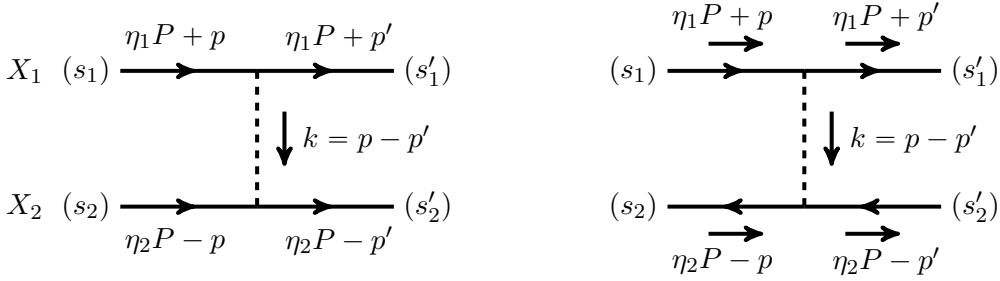

The non-relativistic static potential describing the $X_1-X_2$ interaction is~(see e.g.~\cite{PeskinSchroeder,Petraki:2015hla})
\begin{equation}
V({\bf r}) = -\frac{1}{4 M\mu}\int\frac{d^3k}{(2\pi)^3} 
\mathcal{M}_{\rm 2PI}({\bf k}) \ e^{i{\bf k r}} \,,
\label{eq:V(r)_def}
\end{equation}
where $\mathcal{M}_{\rm 2PI}({\bf k})$ encompasses all 2-particle-irreducible diagrams that contribute to the $X_1-X_2$ elastic scattering. To leading order, this is the one boson exchange shown in \cref{fig:1BED}. 

From the Lagrangians of \cref{eq:L_Scalars_Real,eq:L_Scalars_Complex,eq:L_Fermions}, we find for the interaction of a scalar pair $X_1X_2$, a fermionic pair $X_1X_2$ and a fermion-antifermion pair $X\bar{X}$,
\begin{align}
i\mathcal{M}_{\rm 2PI}^{sc}({\bf k})
&= (-ig_1m_1) (-ig_2m_2) \, \frac{i}{k^2 - m_\varphi^2}
\simeq \frac{ig_1g_2m_1m_2}{ {\bf k}^2 + m_\varphi^2} \,,
\label{eq:MX1X2_scalars} \\
i\mathcal{M}_{\rm 2PI}^f({\bf k})
&= 
[-ig_1 \bar{u}^{s_1'}(\eta_1 P +p') \, u^{s_1}(\eta_1 P + p)]
[-ig_2 \bar{u}^{s_2'}(\eta_2 P -p') \, u^{s_2}(\eta_2 P - p)]
\, \frac{i}{k^2 - m_\varphi^2}
\nonumber \\
&\simeq \frac{i 4g_1g_2m_1m_2}{{\bf k}^2 + m_\varphi^2}\delta^{s_1 s_1'}\delta^{s_2 s_2'} \,, 
\label{eq:MX1X2_fermions}  \\
i\mathcal{M}_{\rm 2PI}^{f, \, X \bar{X}} ({\bf k})
&= (-1)^3
[-ig \bar{u}^{s_1'}(P/2 +p') \, u^{s_1}(P/2 + p)]
[-ig \bar{v}^{s_2} (P/2 -p)  \, v^{s_2'}(P/2 - p')]
\, \frac{i}{k^2 - m_\varphi^2}
\nonumber \\
&\simeq \frac{i 4 g^2 m_X^2}{{\bf k}^2 + m_\varphi^2}\delta^{s_1 s_1'}\delta^{s_2 s_2'} \,, 
\label{eq:MX1X2_fermionantifermion} 
\end{align}
where we made the approximation $k^0 \ll |{\bf k}|$. In \cref{eq:MX1X2_fermionantifermion}, the factor $(-1)^3$ arises from the permutation of the fermion fields in the Wick contractions (see e.g.~\cite[section 4.7]{PeskinSchroeder}). For the spinor contractions in \cref{eq:MX1X2_fermions,eq:MX1X2_fermionantifermion}, we used 
\begin{align}
\bar{u}^{s_j'} (p_j') u^{s_j} (p_j) \simeq \bar{u}^{s_j'} (p_j ) u^{s_j} (p_j) 
&=+2m_j \delta^{s_js_j'}
\label{eq:ubaru}
\\ 
\bar{v}^{s_j}(p_j) v^{s_j'}(p_j') \simeq \bar{v}^{s_j}(p_j) v^{s_j'}(p_j)
&= -2m_j \delta^{s_j s_j'}
\label{eq:vbarv}
\end{align}

From \cref{eq:V(r)_def,eq:MX1X2_scalars,eq:MX1X2_fermions,eq:MX1X2_fermionantifermion}, we find the well known Yukawa potential 
\begin{equation}
V({\bf r})= - \frac{\alpha}{r}e^{-m_\varphi r} \,,
\label{eq:YukawaPotential}
\end{equation}
with $\alpha = \alpha_{sc}$ or $\alpha = \alpha_f$ depending on whether the interacting particles are scalars or fermions respectively, where
\begin{equation}
\alpha_{sc} \equiv \frac{g_1g_2}{16\pi} 
\qquad \text{and} \qquad
\alpha_f \equiv \frac{g_1g_2}{4\pi} \,.
\label{eq:alphas}
\end{equation}
The interaction is attractive if $g_1 g_2 >0$.

%% file: appendix/Wavefunctions.tex
The bound state and scattering state wavefunctions obey the Schr\"odinger equations
\begin{align}
\left[ -\frac{\nabla^2}{2\mu}+V(r)\right] \psi_{n \ell m}({\bf r})
&=-\epsilon_{n\ell}\psi_{\{n\ell m\}}({\bf r}) \,,
\\
\left[ -\frac{\nabla^2}{2\mu}+V(r)\right] \phi_{\bf k}({\bf r})
&=+\epsilon_{\bf k} \phi_{\bf k}({\bf r}) \,.
\end{align}

\paragraph{Coulomb limit, $m_\varphi\rightarrow 0$.}

The ground-state and scattering state wavefunctions are
\begin{align}
\psi^C_{100}({\bf{r}}) 
&= \sqrt{\frac{\kappa^3}{\pi}} \ e^{-\kappa r } \,,
\label{eq:psi_Coul}
\\
\phi^C_{\bf k}({\bf r})
&=\sqrt{S_0^C(\zeta)} 
\ {}_1F_1[i\zeta,1,i(kr-{\bf k r})]e^{i {\bf k r}} \,,
\label{eq:phi_Coul}
\end{align}  
where ${}_1F_1$ is the confluent hypergeometric function of the first kind, and 
\begin{equation}
S_0^C (\zeta) \equiv \frac{2\pi \zeta}{1-e^{-2\pi\zeta}} \,.
\label{eq:S0_Coul}
\end{equation}
$\phi^C_{\bf k}({\bf r})$ can also be decomposed in angular momentum modes as follows
\begin{equation}
\phi^C_{\bf k}({\bf r}) = 
\sqrt{S_0^C(\zeta)} 
\sum_{\ell_S=0}^{\infty}
\frac{\Gamma(1+\ell_S-i\zeta)}{(2\ell_S)! \ \Gamma(1-i\zeta)}
\: (2ikr)^{\ell_S}
\: e^{-ikr}
\: {}_1 F_1 \left(1+\ell_S+i\zeta, \ 2\ell_S+2, \ i 2 k r \right)
\: P_{\ell_S} (\hat{\bf k} \hat{\bf r}) .
\label{eq:phi_Coul_modes}
\end{equation}

\paragraph{Outside the Coulomb regime: Hulthen approximation.} 

For the Yukawa potential \eqref{eq:Yukawa}, the solutions of the Schr\"odinger equation are not known in a closed form. However, it is possible to obtain analytical solutions for the $\ell=0$ modes of the wavefunctions, if we replace the Yukawa for the Hulthen potential~\cite{HulthenA,HulthenB},
\begin{equation}
\label{eq:Hulten}
V_H({\bf r})= - \alpha m_* \frac{e^{-m_* r}}{1-e^{-m_* r}} \,.
\end{equation}
For $m_*\sim m_\varphi$, the Hulthen potential reproduces the behavior of the Yukawa potential at short and large distances. Both potentials admit bound state solutions provided that the screening length scale is sufficiently large. The thresholds for the $n$-th bound level in the Hulthen potential are $m_* \leqslant 2\mu \alpha/n^2$, while for the Yukawa potential \eqref{eq:Yukawa}, bound states exist if $m_\varphi \leqslant \mu \alpha / 0.84$~\cite{Petraki:2016cnz}. We shall pick 
\begin{equation}
m_* = 1.68 m_\varphi \,,
\label{eq:mstar}
\end{equation} 
such that the conditions for the existence of the lowest bound state coincide.

The solutions to the Schr\"odinger equation for the Hulthen potential can be expressed in terms of the two dimensionless parameters
\begin{equation}
\zeta \equiv \alpha / \vrel  
\quad \text{and} \quad
\xi \equiv 2\mu \alpha / m_* \,.
\label{eq:zetaANDxi}	
\end{equation}
Note that $\zeta$ and $\xi$ are always defined using the appropriate fermionic or scalar coupling $\alpha$. 
$\zeta$ compares the average momentum transfer between two unbound particles ($\sim \mu \vrel$) with the relative average momentum of the particles inside the bound state ($\sim \mu \alpha$), while $\xi$ compares the Bohr momentum ($\kappa=\mu\alpha$) that determines the size of the bound state, with the screening scale ($m_*$) that determines the range of the interaction. The interaction manifests as long-range roughly if $\xi \gtrsim 1$; this is the regime where non-perturbative phenomena, the Sommerfeld effect and bound states, emerge. The Coulomb limit is recovered for $\xi\rightarrow\infty$.

For the ground state, the binding energy is
\begin{equation}
\epsilon_{10} = \frac{\kappa^2 (1-1/\xi)^2}{2\mu} \,,
\label{eq:Hulthen_epsilon10}
\end{equation}
and the wavefunction reads
\begin{align}
\psi_{100}^{} ({\bf{r}}) 
&=\xi \, \sqrt{\frac{\kappa^3}{\pi} \left(1-\frac{1}{\xi^2}\right)} 
\ \frac{\sinh(\kappa r/\xi)\exp(-\kappa r)}{\kappa r} \,.
\label{eq:Hulthen_psi}
\end{align}        

For the scattering state, we make the partial-wave decomposition
\begin{equation}
\phi_{\bf k}({\bf r}) = \sum_{\ell_S=0}^{\infty}  
w_{\ell_S}^{} (m_* r) \ P_{\ell_S}({\bf\hat{k} \cdot \hat{r}}) \,.
\label{eq:phi_decomposition}
\end{equation}
It is possible to find an analytical solution for $\ell_S = 0$ only, which suffices for our purposes,
\begin{equation}
w_0 (z) =  
\sqrt{S_0(\zeta,\xi)}
\ e^{-i \xi z/(2\zeta)} 
\ \left(\frac{1-e^{-z}}{z}\right) 
\ {}_2F_1(a_0,b_0,2; \ 1-e^{-z}) \,,
\label{eq:Hulthen_w0}
\end{equation}
where ${}_2F_1$ is the hypergeometric function and
\begin{align}
a_0 &\equiv 1+\frac{i\xi}{2\zeta}(1-\sqrt{1-4\zeta^2/\xi}) \,,
\label{eq:Hulthen_a0}
\\
b_0 &\equiv 1+\frac{i\xi}{2\zeta}(1+\sqrt{1-4\zeta^2/\xi}) \,,
\label{eq:Hulthen_b0}
\\
\sqrt{S_0(\zeta,\xi)} 
&\equiv \frac{i\zeta}{\xi}\left[\frac{\Gamma(a_0)\Gamma(b_0)  }{\Gamma(i\xi/\zeta)}\right]^* \,.
\label{eq:Hulthen_S0sqrt}
\end{align}
From \cref{eq:Hulthen_S0sqrt}, we obtain 
\begin{equation}
S_0(\zeta,\xi) =
\frac{2\pi \zeta \sinh (\pi \xi/\zeta)}
{\cosh(\pi\xi/\zeta) - \cosh [(\pi\xi/\zeta)\sqrt{1-4\zeta^2/\xi} ]} \,,
\label{eq:Hulthen_S0}
\end{equation}
which reduces to \cref{eq:S0_Coul} in the limit $\xi \to \infty$.

%% file: appendix/OverlapIntegral.tex
For the capture into bound states via emission of two scalars, we need to compute the overlap integrals
\begin{align}
{\cal V}_{{\bf k},\{n\ell m\}} 
&\equiv (8\pi \kappa)^{1/2} 
\int \frac{d^3 q}{(2\pi)^3} \frac{d^3 p}{(2\pi)^3}
\frac
{\tilde{\phi}_{\bf k} ({\bf q}) \tilde{\psi}_{n\ell m}^* ({\bf p})}
{({\bf q}-{\bf p})^2 + m_\varphi^2}  \,,
\label{eq:Overlap_V_def_re}
\\
{\cal R}_{{\bf k},\{n\ell m\}} 
&\equiv (8 \pi \kappa^5)^{1/2}
\int \frac{d^3 q}{(2\pi)^3} \frac{d^3 p}{(2\pi)^3}
\frac
{\tilde{\phi}_{\bf k} ({\bf q}) \tilde{\psi}_{n\ell m}^* ({\bf p})}
{[({\bf q}-{\bf p})^2 + m_\varphi^2]^2} \,.
\label{eq:Overlap_R_def_re}
\end{align} 
We Fourier transform the wavefunctions as follows
\begin{align}
\psi_{n\ell m}({\bf r})
=\int \frac{d^3 p}{(2\pi)^3 } \ \tilde{\psi}_{\{n \ell m\}}({\bf p}) \, e^{i {\bf p r}}, 
&\qquad
\tilde{\psi}_{n\ell m}({\bf p})
=\int d^3r \ \psi_{n\ell m}({\bf r})e^{-i {\bf pr}} \,,
\label{eq:FT_psi}
\\
\phi_{\bf k}({\bf r}) 
= \int \frac{d^3 p}{(2\pi)^3 } \ \tilde{\phi}_{\bf k}({\bf p}) \, e^{i {\bf p r}}, 
&\qquad 
\tilde{\phi}_{\bf k}({\bf p})
=\int d^3r \ \phi_{\bf k}({\bf r})e^{-i {\bf pr}} \,.
\label{eq:FT_phi}
\end{align}
The overlap integrals become
\begin{align}
{\cal V}_{{\bf k},\{n\ell m\}} 
&= \left(\frac{\kappa}{2 \pi} \right)^{1/2} 
\int d^3 r \, 
\phi_{\bf k} ({\bf r}) \psi_{n\ell m}^* ({\bf r})
\frac{e^{-m_\varphi r}}{r} ,
\label{eq:Overlap_V_FT}
\\
{\cal R}_{{\bf k},\{n\ell m\}} 
&= \left(\frac{\kappa^5 }{8\pi}\right)^{1/2} \frac{1}{m_\varphi}
\int d^3 r \, 
\phi_{\bf k} ({\bf r}) \psi_{n\ell m}^* ({\bf r}) 
\, e^{-m_\varphi r}
\simeq 
-\left(\frac{\kappa^5 }{8\pi}\right)^{1/2}
\int d^3 r \, 
\phi_{\bf k} ({\bf r}) \psi_{n\ell m}^* ({\bf r}) 
\, r .
\label{eq:Overlap_R_FT}
\end{align} 
In the second step in \cref{eq:Overlap_R_FT}, we expanded the decaying exponential inside the integral. The bound-state wavefunction implies that the integrand is significant for $r \lesssim n/(\mu \alpha)$ while the kinematic threshold~\eqref{eq:BSF1_KinematicThresshold} for BSF imposes $m_\varphi < \mu \alpha^2/(2n^2)$; therefore $m_\varphi r \lesssim \alpha \ll 1$.  
The zeroth order term  in the expansion vanishes due to the orthogonality of the wavefunctions, leaving the first order term to be the dominant contribution. 
In the following, we will focus on the capture into the ground state, $\{n\ell m\} = \{100\}$.

While in the Coulomb limit ($m_\varphi \to 0$) it is possible to obtain analytical expressions for the integrals \cref{eq:Overlap_V_FT,eq:Overlap_R_FT}, outside the Coulomb regime they should be evaluated numerically. However, it is possible to obtain an analytical approximation, using the Hulthen potential (cf.~\cref{App:wf}). Since $\xi \equiv 2\mu\alpha/m_* \gg 1$ in all the parameter space where BSF is kinematically allowed to occur [cf.~\cref{eq:BSF1_KinematicThresshold}], the bound state wavefunction $\psi_{100} ({\bf r})$ can be well approximated by its Coulomb value.  On the other hand, the scattering-state wavefunction $\phi_{\bf k} ({\bf r})$ is close to its Coulomb  limit for $r \lesssim 1/m_*$, up to an overall normalisation which, for the $\ell_S=0$ mode, is determined by the factor $S_0(\zeta,\xi)$ of \cref{eq:Hulthen_S0}. Since $\xi \gg 1$, this encompasses all the range in which $\psi_{100}({\bf r})$ and therefore the integrands in \cref{eq:Overlap_R_FT,eq:Overlap_V_FT} are important, $r \lesssim 1/(\mu \alpha)$.  It follows that ${\cal V}_{{\bf k},\{100\}}$ and ${\cal R}_{{\bf k},\{100\}}$ -- which depend only on the $\ell_S=0$ mode of the scattering state wavefunction -- can be well approximated by their Coulomb values even outside the Coulomb regime provided that we replace $S_0^C (\zeta)$ in \cref{eq:phi_Coul} with $S_0(\zeta, \xi)$.

Following ref.~\cite{Pearce:2013ola} (see also~\cite{Petraki:2015hla,Petraki:2016cnz,Harz:2018csl}), we shall use the identity~\cite{AkhiezerMerenkov_sigmaHydrogen}
\begin{equation}
\int d^3r \frac{e^{i{(\bf k-\Gamma)r}-\kappa r}}{4\pi r} 
{_1}F_1[i\zeta,1,i(kr-{\bf kr})]
= \frac{[{\bf \Gamma}^2+(\kappa-ik)^2]^{-i\zeta}}{{[{\bf (k-\Gamma)^2}+\kappa^2]^{1-i\zeta}}}
\equiv f_{{\bf k,\Gamma}}(\kappa) \,.
\label{eq:f}
\end{equation}
Then, \cref{eq:Overlap_V_FT,eq:Overlap_R_FT} become
\begin{align}
{\cal V}_{{\bf k},\{100\}}  &\simeq 
+\sqrt{8\kappa^4 S_0(\zeta, \xi)} \ f_{\bf k, \Gamma=0} (\kappa) \,,
\label{eq:Vwrtf}
\\
{\cal R}_{{\bf k},\{100\}}  &\simeq 
-\sqrt{ 2 \kappa^8 S_0(\zeta,\xi) } \
\left[ 
\frac{\partial^2 f_{{\bf k,\Gamma=0}}(\kappa)}{\partial \kappa^2}
\right] \,.
\label{eq:Rwrtf}
\end{align}
Taking into account that $\kappa / k = \zeta$, we arrive at
\begin{align}
{\cal V}_{{\bf k},\{100\}}  &\simeq 
\sqrt{8 S_0(\zeta, \xi)} 
\left(\frac{\zeta^2}{1+\zeta^2} \right) 
\ e^{-2\zeta \, {\rm arccot} \, \zeta} \,,
\label{eq:V_GroundState}
\\
{\cal R}_{{\bf k},\{100\}}  &\simeq 
\sqrt{8 S_0(\zeta, \xi)} 
\left(\frac{\zeta^2}{1+\zeta^2} \right)^2 
\ e^{-2\zeta \, {\rm arccot} \, \zeta}  \,.
\label{eq:R_GroundState}
\end{align}

%% file: BSF_ScalarMed_arXiv_v3.bbl
\providecommand{\href}[2]{#2}\begingroup\raggedright\begin{thebibliography}{10}

\bibitem{Tulin:2017ara}
S.~Tulin and H.-B. Yu, \emph{{Dark Matter Self-interactions and Small Scale
  Structure}}, \href{https://doi.org/10.1016/j.physrep.2017.11.004}{\emph{Phys.
  Rept.} {\bfseries 730} (2018) 1}
  [\href{https://arxiv.org/abs/1705.02358}{{\ttfamily 1705.02358}}].

\bibitem{Sommerfeld:1931}
A.~Sommerfeld, \emph{{{\"U}ber die Beugung und Bremsung der Elektronen}},
  {\emph{Ann. Phys.} {\bfseries 403} (1931) 257}.

\bibitem{Sakharov:1948yq}
A.~D. Sakharov, \emph{{Interaction of an Electron and Positron in Pair
  Production}},
  \href{https://doi.org/10.1070/PU1991v034n05ABEH002492}{\emph{Zh. Eksp. Teor.
  Fiz.} {\bfseries 18} (1948) 631}.

\bibitem{Harz:2017dlj}
J.~Harz and K.~Petraki, \emph{{Higgs Enhancement for the Dark Matter Relic
  Density}}, \href{https://doi.org/10.1103/PhysRevD.97.075041}{\emph{Phys.
  Rev.} {\bfseries D97} (2018) 075041}
  [\href{https://arxiv.org/abs/1711.03552}{{\ttfamily 1711.03552}}].

\bibitem{vonHarling:2014kha}
B.~von Harling and K.~Petraki, \emph{{Bound-state formation for thermal relic
  dark matter and unitarity}},
  \href{https://doi.org/10.1088/1475-7516/2014/12/033}{\emph{JCAP} {\bfseries
  12} (2014) 033} [\href{https://arxiv.org/abs/1407.7874}{{\ttfamily
  1407.7874}}].

\bibitem{Pospelov:2008jd}
M.~Pospelov and A.~Ritz, \emph{{Astrophysical Signatures of Secluded Dark
  Matter}},
  \href{https://doi.org/10.1016/j.physletb.2008.12.012}{\emph{Phys.Lett.}
  {\bfseries B671} (2009) 391}
  [\href{https://arxiv.org/abs/0810.1502}{{\ttfamily 0810.1502}}].

\bibitem{An:2016gad}
H.~An, M.~B. Wise and Y.~Zhang, \emph{{Effects of Bound States on Dark Matter
  Annihilation}}, \href{https://doi.org/10.1103/PhysRevD.93.115020}{\emph{Phys.
  Rev.} {\bfseries D93} (2016) 115020}
  [\href{https://arxiv.org/abs/1604.01776}{{\ttfamily 1604.01776}}].

\bibitem{Asadi:2016ybp}
P.~Asadi, M.~Baumgart, P.~J. Fitzpatrick, E.~Krupczak and T.~R. Slatyer,
  \emph{{Capture and Decay of Electroweak WIMPonium}},
  \href{https://doi.org/10.1088/1475-7516/2017/02/005}{\emph{JCAP} {\bfseries
  1702} (2017) 005} [\href{https://arxiv.org/abs/1610.07617}{{\ttfamily
  1610.07617}}].

\bibitem{Petraki:2016cnz}
K.~Petraki, M.~Postma and J.~de~Vries, \emph{{Radiative bound-state-formation
  cross-sections for dark matter interacting via a Yukawa potential}},
  \href{https://doi.org/10.1007/JHEP04(2017)077}{\emph{JHEP} {\bfseries 04}
  (2017) 077} [\href{https://arxiv.org/abs/1611.01394}{{\ttfamily
  1611.01394}}].

\bibitem{Cirelli:2016rnw}
M.~Cirelli, P.~Panci, K.~Petraki, F.~Sala and M.~Taoso, \emph{{Dark Matter's
  secret liaisons: phenomenology of a dark U(1) sector with bound states}},
  \href{https://doi.org/10.1088/1475-7516/2017/05/036}{\emph{JCAP} {\bfseries
  1705} (2017) 036} [\href{https://arxiv.org/abs/1612.07295}{{\ttfamily
  1612.07295}}].

\bibitem{Kouvaris:2016ltf}
C.~Kouvaris, K.~Langaeble and N.~G. Nielsen, \emph{{The Spectrum of Darkonium
  in the Sun}},
  \href{https://doi.org/10.1088/1475-7516/2016/10/012}{\emph{JCAP} {\bfseries
  1610} (2016) 012} [\href{https://arxiv.org/abs/1607.00374}{{\ttfamily
  1607.00374}}].

\bibitem{Baldes:2017gzw}
I.~Baldes and K.~Petraki, \emph{{Asymmetric thermal-relic dark matter:
  Sommerfeld-enhanced freeze-out, annihilation signals and unitarity bounds}},
  \href{https://doi.org/10.1088/1475-7516/2017/09/028}{\emph{JCAP} {\bfseries
  1709} (2017) 028} [\href{https://arxiv.org/abs/1703.00478}{{\ttfamily
  1703.00478}}].

\bibitem{Baldes:2017gzu}
I.~Baldes, M.~Cirelli, P.~Panci, K.~Petraki, F.~Sala and M.~Taoso,
  \emph{{Asymmetric dark matter: residual annihilations and
  self-interactions}},
  \href{https://doi.org/10.21468/SciPostPhys.4.6.041}{\emph{SciPost Phys.}
  {\bfseries 4} (2018) 041} [\href{https://arxiv.org/abs/1712.07489}{{\ttfamily
  1712.07489}}].

\bibitem{MarchRussell:2008tu}
J.~D. March-Russell and S.~M. West, \emph{{WIMPonium and Boost Factors for
  Indirect Dark Matter Detection}},
  \href{https://doi.org/10.1016/j.physletb.2009.04.010}{\emph{Phys.Lett.}
  {\bfseries B676} (2009) 133}
  [\href{https://arxiv.org/abs/0812.0559}{{\ttfamily 0812.0559}}].

\bibitem{An:2016kie}
H.~An, M.~B. Wise and Y.~Zhang, \emph{{Strong CMB Constraint On P-Wave
  Annihilating Dark Matter}},
  \href{https://doi.org/10.1016/j.physletb.2017.08.010}{\emph{Phys. Lett.}
  {\bfseries B773} (2017) 121}
  [\href{https://arxiv.org/abs/1606.02305}{{\ttfamily 1606.02305}}].

\bibitem{Petraki:2014uza}
K.~Petraki, L.~Pearce and A.~Kusenko, \emph{{Self-interacting asymmetric dark
  matter coupled to a light massive dark photon}},
  \href{https://doi.org/10.1088/1475-7516/2014/07/039}{\emph{JCAP} {\bfseries
  1407} (2014) 039} [\href{https://arxiv.org/abs/1403.1077}{{\ttfamily
  1403.1077}}].

\bibitem{Pearce:2015zca}
L.~Pearce, K.~Petraki and A.~Kusenko, \emph{{Signals from dark atom formation
  in halos}},
  \href{https://doi.org/10.1103/PhysRevD.91.083532}{\emph{Phys.Rev.} {\bfseries
  D91} (2015) 083532} [\href{https://arxiv.org/abs/1502.01755}{{\ttfamily
  1502.01755}}].

\bibitem{Cline:2014eaa}
J.~M. Cline, Y.~Farzan, Z.~Liu, G.~D. Moore and W.~Xue, \emph{{3.5 keV x rays
  as the “21 cm line” of dark atoms, and a link to light sterile
  neutrinos}},
  \href{https://doi.org/10.1103/PhysRevD.89.121302}{\emph{Phys.Rev.} {\bfseries
  D89} (2014) 121302} [\href{https://arxiv.org/abs/1404.3729}{{\ttfamily
  1404.3729}}].

\bibitem{Pearce:2013ola}
L.~Pearce and A.~Kusenko, \emph{{Indirect Detection of Self-Interacting
  Asymmetric Dark Matter}},
  \href{https://doi.org/10.1103/PhysRevD.87.123531}{\emph{Phys.Rev.} {\bfseries
  D87} (2013) 123531} [\href{https://arxiv.org/abs/1303.7294}{{\ttfamily
  1303.7294}}].

\bibitem{Detmold:2014qqa}
W.~Detmold, M.~McCullough and A.~Pochinsky, \emph{{Dark Nuclei I: Cosmology and
  Indirect Detection}},
  \href{https://doi.org/10.1103/PhysRevD.90.115013}{\emph{Phys.Rev.} {\bfseries
  D90} (2014) 115013} [\href{https://arxiv.org/abs/1406.2276}{{\ttfamily
  1406.2276}}].

\bibitem{Laha:2013gva}
R.~Laha and E.~Braaten, \emph{{Direct detection of dark matter in universal
  bound states}},
  \href{https://doi.org/10.1103/PhysRevD.89.103510}{\emph{Phys.Rev.} {\bfseries
  D89} (2014) 103510} [\href{https://arxiv.org/abs/1311.6386}{{\ttfamily
  1311.6386}}].

\bibitem{Butcher:2016hic}
A.~Butcher, R.~Kirk, J.~Monroe and S.~M. West, \emph{{Can Tonne-Scale Direct
  Detection Experiments Discover Nuclear Dark Matter?}},
  \href{https://doi.org/10.1088/1475-7516/2017/10/035}{\emph{JCAP} {\bfseries
  1710} (2017) 035} [\href{https://arxiv.org/abs/1610.01840}{{\ttfamily
  1610.01840}}].

\bibitem{Krnjaic:2014xza}
G.~Krnjaic and K.~Sigurdson, \emph{{Big Bang Darkleosynthesis}},
  \href{https://doi.org/10.1016/j.physletb.2015.11.001}{\emph{Phys. Lett.}
  {\bfseries B751} (2015) 464}
  [\href{https://arxiv.org/abs/1406.1171}{{\ttfamily 1406.1171}}].

\bibitem{Wise:2014jva}
M.~B. Wise and Y.~Zhang, \emph{{Stable Bound States of Asymmetric Dark
  Matter}}, \href{https://doi.org/10.1103/PhysRevD.90.055030}{\emph{Phys.Rev.}
  {\bfseries D90} (2014) 055030}
  [\href{https://arxiv.org/abs/1407.4121}{{\ttfamily 1407.4121}}].

\bibitem{Wise:2014ola}
M.~B. Wise and Y.~Zhang, \emph{{Yukawa Bound States of a Large Number of
  Fermions}}, \href{https://doi.org/10.1007/JHEP02(2015)023}{\emph{JHEP}
  {\bfseries 02} (2015) 023} [\href{https://arxiv.org/abs/1411.1772}{{\ttfamily
  1411.1772}}].

\bibitem{Coleman:1985ki}
S.~R. Coleman, \emph{{Q Balls}},
  \href{https://doi.org/10.1016/0550-3213(85)90286-X;
  10.1016/0550-3213(85)90286-X}{\emph{Nucl.Phys.} {\bfseries B262} (1985) 263}.

\bibitem{Kusenko:1997ad}
A.~Kusenko, \emph{{Small Q balls}},
  \href{https://doi.org/10.1016/S0370-2693(97)00582-0}{\emph{Phys.Lett.}
  {\bfseries B404} (1997) 285}
  [\href{https://arxiv.org/abs/hep-th/9704073}{{\ttfamily hep-th/9704073}}].

\bibitem{Kusenko:1997zq}
A.~Kusenko, \emph{{Solitons in the supersymmetric extensions of the standard
  model}},
  \href{https://doi.org/10.1016/S0370-2693(97)00584-4}{\emph{Phys.Lett.}
  {\bfseries B405} (1997) 108}
  [\href{https://arxiv.org/abs/hep-ph/9704273}{{\ttfamily hep-ph/9704273}}].

\bibitem{Kusenko:1997si}
A.~Kusenko and M.~E. Shaposhnikov, \emph{{Supersymmetric Q balls as dark
  matter}},
  \href{https://doi.org/10.1016/S0370-2693(97)01375-0}{\emph{Phys.Lett.}
  {\bfseries B418} (1998) 46}
  [\href{https://arxiv.org/abs/hep-ph/9709492}{{\ttfamily hep-ph/9709492}}].

\bibitem{Shepherd:2009sa}
W.~Shepherd, T.~M. Tait and G.~Zaharijas, \emph{{Bound states of weakly
  interacting dark matter}},
  \href{https://doi.org/10.1103/PhysRevD.79.055022}{\emph{Phys.Rev.} {\bfseries
  D79} (2009) 055022} [\href{https://arxiv.org/abs/0901.2125}{{\ttfamily
  0901.2125}}].

\bibitem{Petraki:2015hla}
K.~Petraki, M.~Postma and M.~Wiechers, \emph{{Dark-matter bound states from
  Feynman diagrams}},
  \href{https://doi.org/10.1007/JHEP06(2015)128}{\emph{JHEP} {\bfseries 1506}
  (2015) 128} [\href{https://arxiv.org/abs/1505.00109}{{\ttfamily
  1505.00109}}].

\bibitem{Kim:2016kxt}
S.~Kim and M.~Laine, \emph{{On thermal corrections to near-threshold
  annihilation}},
  \href{https://doi.org/10.1088/1475-7516/2017/01/013}{\emph{JCAP} {\bfseries
  1701} (2017) 013} [\href{https://arxiv.org/abs/1609.00474}{{\ttfamily
  1609.00474}}].

\bibitem{Kim:2016zyy}
S.~Kim and M.~Laine, \emph{{Rapid thermal co-annihilation through bound states
  in QCD}}, \href{https://doi.org/10.1007/JHEP07(2016)143}{\emph{JHEP}
  {\bfseries 07} (2016) 143}
  [\href{https://arxiv.org/abs/1602.08105}{{\ttfamily 1602.08105}}].

\bibitem{Harz:2018csl}
J.~Harz and K.~Petraki, \emph{{Radiative bound-state formation in unbroken
  perturbative non-Abelian theories and implications for dark matter}},
  \href{https://arxiv.org/abs/1805.01200}{{\ttfamily 1805.01200}}.

\bibitem{Geller:2018biy}
M.~Geller, S.~Iwamoto, G.~Lee, Y.~Shadmi and O.~Telem, \emph{{Dark quarkonium
  formation in the early universe}},
  \href{https://doi.org/10.1007/JHEP06(2018)135}{\emph{JHEP} {\bfseries 06}
  (2018) 135} [\href{https://arxiv.org/abs/1802.07720}{{\ttfamily
  1802.07720}}].

\bibitem{Biondini:2018pwp}
S.~Biondini and M.~Laine, \emph{{Thermal dark matter co-annihilating with a
  strongly interacting scalar}},
  \href{https://doi.org/10.1007/JHEP04(2018)072}{\emph{JHEP} {\bfseries 04}
  (2018) 072} [\href{https://arxiv.org/abs/1801.05821}{{\ttfamily
  1801.05821}}].

\bibitem{Biondini:2017ufr}
S.~Biondini and M.~Laine, \emph{{Re-derived overclosure bound for the inert
  doublet model}}, \href{https://doi.org/10.1007/JHEP08(2017)047}{\emph{JHEP}
  {\bfseries 08} (2017) 047}
  [\href{https://arxiv.org/abs/1706.01894}{{\ttfamily 1706.01894}}].

\bibitem{Kaplan:2009de}
D.~E. Kaplan, G.~Z. Krnjaic, K.~R. Rehermann and C.~M. Wells, \emph{{Atomic
  Dark Matter}}, \href{https://doi.org/10.1088/1475-7516}{\emph{JCAP}
  {\bfseries 1005} (2010) 021}
  [\href{https://arxiv.org/abs/0909.0753}{{\ttfamily 0909.0753}}].

\bibitem{CyrRacine:2012fz}
F.-Y. Cyr-Racine and K.~Sigurdson, \emph{{The Cosmology of Atomic Dark
  Matter}}, \href{https://doi.org/10.1103/PhysRevD.87.103515}{\emph{Phys.Rev.}
  {\bfseries D87} (2013) 103515}
  [\href{https://arxiv.org/abs/1209.5752}{{\ttfamily 1209.5752}}].

\bibitem{Lonsdale:2014wwa}
S.~J. Lonsdale and R.~R. Volkas, \emph{{Grand unified hidden-sector dark
  matter}}, \href{https://doi.org/10.1103/PhysRevD.90.083501,
  10.1103/PhysRevD.91.129906}{\emph{Phys. Rev.} {\bfseries D90} (2014) 083501}
  [\href{https://arxiv.org/abs/1407.4192}{{\ttfamily 1407.4192}}].

\bibitem{Boddy:2014yra}
K.~K. Boddy, J.~L. Feng, M.~Kaplinghat and T.~M.~P. Tait,
  \emph{{Self-Interacting Dark Matter from a Non-Abelian Hidden Sector}},
  \href{https://doi.org/10.1103/PhysRevD.89.115017}{\emph{Phys.Rev.} {\bfseries
  D89} (2014) 115017} [\href{https://arxiv.org/abs/1402.3629}{{\ttfamily
  1402.3629}}].

\bibitem{Kribs:2016cew}
G.~D. Kribs and E.~T. Neil, \emph{{Review of strongly-coupled composite dark
  matter models and lattice simulations}},
  \href{https://doi.org/10.1142/S0217751X16430041}{\emph{Int. J. Mod. Phys.}
  {\bfseries A31} (2016) 1643004}
  [\href{https://arxiv.org/abs/1604.04627}{{\ttfamily 1604.04627}}].

\bibitem{Lonsdale:2017mzg}
S.~J. Lonsdale, M.~Schroor and R.~R. Volkas, \emph{{Asymmetric Dark Matter and
  the hadronic spectra of hidden QCD}},
  \href{https://doi.org/10.1103/PhysRevD.96.055027}{\emph{Phys. Rev.}
  {\bfseries D96} (2017) 055027}
  [\href{https://arxiv.org/abs/1704.05213}{{\ttfamily 1704.05213}}].

\bibitem{Lonsdale:2018xwd}
S.~J. Lonsdale and R.~R. Volkas, \emph{{Comprehensive asymmetric dark matter
  model}}, \href{https://doi.org/10.1103/PhysRevD.97.103510}{\emph{Phys. Rev.}
  {\bfseries D97} (2018) 103510}
  [\href{https://arxiv.org/abs/1801.05561}{{\ttfamily 1801.05561}}].

\bibitem{Gresham:2017cvl}
M.~I. Gresham, H.~K. Lou and K.~M. Zurek, \emph{{Early Universe synthesis of
  asymmetric dark matter nuggets}},
  \href{https://doi.org/10.1103/PhysRevD.97.036003}{\emph{Phys. Rev.}
  {\bfseries D97} (2018) 036003}
  [\href{https://arxiv.org/abs/1707.02316}{{\ttfamily 1707.02316}}].

\bibitem{Braaten:2018xuw}
E.~Braaten, D.~Kang and R.~Laha, \emph{{Production of dark-matter bound states
  in the early universe by three-body recombination}},
  \href{https://arxiv.org/abs/1806.00609}{{\ttfamily 1806.00609}}.

\bibitem{Itzykson:1980rh}
C.~Itzykson and J.~Zuber, \emph{{Quantum field theory}}. 1980.

\bibitem{Pineda:1997bj}
A.~Pineda and J.~Soto, \emph{{Effective field theory for ultrasoft momenta in
  NRQCD and NRQED}},
  \href{https://doi.org/10.1016/S0920-5632(97)01102-X}{\emph{Nucl. Phys. Proc.
  Suppl.} {\bfseries 64} (1998) 428}
  [\href{https://arxiv.org/abs/hep-ph/9707481}{{\ttfamily hep-ph/9707481}}].

\bibitem{Brambilla:1999xf}
N.~Brambilla, A.~Pineda, J.~Soto and A.~Vairo, \emph{{Potential NRQCD: An
  Effective theory for heavy quarkonium}},
  \href{https://doi.org/10.1016/S0550-3213(99)00693-8}{\emph{Nucl. Phys.}
  {\bfseries B566} (2000) 275}
  [\href{https://arxiv.org/abs/hep-ph/9907240}{{\ttfamily hep-ph/9907240}}].

\bibitem{Beneke:1999zr}
M.~Beneke, \emph{{Perturbative heavy quark - anti-quark systems}},
  \href{https://arxiv.org/abs/hep-ph/9911490}{{\ttfamily hep-ph/9911490}}.

\bibitem{Manohar:1999xd}
A.~V. Manohar and I.~W. Stewart, \emph{{Renormalization group analysis of the
  QCD quark potential to order v**2}},
  \href{https://doi.org/10.1103/PhysRevD.62.014033}{\emph{Phys. Rev.}
  {\bfseries D62} (2000) 014033}
  [\href{https://arxiv.org/abs/hep-ph/9912226}{{\ttfamily hep-ph/9912226}}].

\bibitem{Manohar:2000kr}
A.~V. Manohar and I.~W. Stewart, \emph{{Running of the heavy quark production
  current and 1 / v potential in QCD}},
  \href{https://doi.org/10.1103/PhysRevD.63.054004}{\emph{Phys. Rev.}
  {\bfseries D63} (2001) 054004}
  [\href{https://arxiv.org/abs/hep-ph/0003107}{{\ttfamily hep-ph/0003107}}].

\bibitem{Pineda:2001ra}
A.~Pineda, \emph{{Renormalization group improvement of the NRQCD Lagrangian and
  heavy quarkonium spectrum}},
  \href{https://doi.org/10.1103/PhysRevD.65.074007}{\emph{Phys. Rev.}
  {\bfseries D65} (2002) 074007}
  [\href{https://arxiv.org/abs/hep-ph/0109117}{{\ttfamily hep-ph/0109117}}].

\bibitem{Brambilla:2004jw}
N.~Brambilla, A.~Pineda, J.~Soto and A.~Vairo, \emph{{Effective field theories
  for heavy quarkonium}},
  \href{https://doi.org/10.1103/RevModPhys.77.1423}{\emph{Rev. Mod. Phys.}
  {\bfseries 77} (2005) 1423}
  [\href{https://arxiv.org/abs/hep-ph/0410047}{{\ttfamily hep-ph/0410047}}].

\bibitem{Pineda:2011aw}
A.~Pineda, \emph{{Next-to-leading ultrasoft running of the heavy quarkonium
  potentials and spectrum: Spin-independent case}},
  \href{https://doi.org/10.1103/PhysRevD.84.014012}{\emph{Phys. Rev.}
  {\bfseries D84} (2011) 014012}
  [\href{https://arxiv.org/abs/1101.3269}{{\ttfamily 1101.3269}}].

\bibitem{Hoang:2011gy}
A.~H. Hoang and M.~Stahlhofen, \emph{{Ultrasoft NLL Running of the
  Nonrelativistic O(v) QCD Quark Potential}},
  \href{https://doi.org/10.1007/JHEP06(2011)088}{\emph{JHEP} {\bfseries 06}
  (2011) 088} [\href{https://arxiv.org/abs/1102.0269}{{\ttfamily 1102.0269}}].

\bibitem{Braaten:2017kci}
E.~Braaten, E.~Johnson and H.~Zhang, \emph{{Zero-range effective field theory
  for resonant wino dark matter. Part II. Coulomb resummation}},
  \href{https://doi.org/10.1007/JHEP02(2018)150}{\emph{JHEP} {\bfseries 02}
  (2018) 150} [\href{https://arxiv.org/abs/1708.07155}{{\ttfamily
  1708.07155}}].

\bibitem{Braaten:2017gpq}
E.~Braaten, E.~Johnson and H.~Zhang, \emph{{Zero-range effective field theory
  for resonant wino dark matter. Part I. Framework}},
  \href{https://doi.org/10.1007/JHEP11(2017)108}{\emph{JHEP} {\bfseries 11}
  (2017) 108} [\href{https://arxiv.org/abs/1706.02253}{{\ttfamily
  1706.02253}}].

\bibitem{Braaten:2017dwq}
E.~Braaten, E.~Johnson and H.~Zhang, \emph{{Zero-range effective field theory
  for resonant wino dark matter. Part III. Annihilation effects}},
  \href{https://doi.org/10.1007/JHEP05(2018)062}{\emph{JHEP} {\bfseries 05}
  (2018) 062} [\href{https://arxiv.org/abs/1712.07142}{{\ttfamily
  1712.07142}}].

\bibitem{Belotsky:2015osa}
K.~M. Belotsky, E.~A. Esipova and A.~A. Kirillov, \emph{{On the classical
  description of the recombination of dark matter particles with a Coulomb-like
  interaction}},
  \href{https://doi.org/10.1016/j.physletb.2016.08.009}{\emph{Phys. Lett.}
  {\bfseries B761} (2016) 81}
  [\href{https://arxiv.org/abs/1506.03094}{{\ttfamily 1506.03094}}].

\bibitem{Bringmann:2016din}
T.~Bringmann, F.~Kahlhoefer, K.~Schmidt-Hoberg and P.~Walia, \emph{{Strong
  constraints on self-interacting dark matter with light mediators}},
  \href{https://arxiv.org/abs/1612.00845}{{\ttfamily 1612.00845}}.

\bibitem{Kahlhoefer:2017umn}
F.~Kahlhoefer, K.~Schmidt-Hoberg and S.~Wild, \emph{{Dark matter
  self-interactions from a general spin-0 mediator}},
  \href{https://doi.org/10.1088/1475-7516/2017/08/003}{\emph{JCAP} {\bfseries
  1708} (2017) 003} [\href{https://arxiv.org/abs/1704.02149}{{\ttfamily
  1704.02149}}].

\bibitem{Bell:2017irk}
N.~F. Bell, Y.~Cai, J.~B. Dent, R.~K. Leane and T.~J. Weiler, \emph{{Enhancing
  Dark Matter Annihilation Rates with Dark Bremsstrahlung}},
  \href{https://doi.org/10.1103/PhysRevD.96.023011}{\emph{Phys. Rev.}
  {\bfseries D96} (2017) 023011}
  [\href{https://arxiv.org/abs/1705.01105}{{\ttfamily 1705.01105}}].

\bibitem{Binder:2017lkj}
T.~Binder, M.~Gustafsson, A.~Kamada, S.~M.~R. Sandner and M.~Wiesner,
  \emph{{Reannihilation of self-interacting dark matter}},
  \href{https://doi.org/10.1103/PhysRevD.97.123004}{\emph{Phys. Rev.}
  {\bfseries D97} (2018) 123004}
  [\href{https://arxiv.org/abs/1712.01246}{{\ttfamily 1712.01246}}].

\bibitem{Petraki:2013wwa}
K.~Petraki and R.~R. Volkas, \emph{{Review of asymmetric dark matter}},
  \href{https://doi.org/10.1142/S0217751X13300287}{\emph{Int.J.Mod.Phys.}
  {\bfseries A28} (2013) 1330028}
  [\href{https://arxiv.org/abs/1305.4939}{{\ttfamily 1305.4939}}].

\bibitem{Kusenko:1997hj}
A.~Kusenko, \emph{{Phase transitions precipitated by solitosynthesis}},
  \href{https://doi.org/10.1016/S0370-2693(97)00700-4}{\emph{Phys.Lett.}
  {\bfseries B406} (1997) 26}
  [\href{https://arxiv.org/abs/hep-ph/9705361}{{\ttfamily hep-ph/9705361}}].

\bibitem{Postma:2001ea}
M.~Postma, \emph{{Solitosynthesis of Q balls}},
  \href{https://doi.org/10.1103/PhysRevD.65.085035}{\emph{Phys.Rev.} {\bfseries
  D65} (2002) 085035} [\href{https://arxiv.org/abs/hep-ph/0110199}{{\ttfamily
  hep-ph/0110199}}].

\bibitem{Pearce:2012jp}
L.~Pearce, \emph{{Solitosynthesis induced phase transitions}},
  \href{https://doi.org/10.1103/PhysRevD.85.125022}{\emph{Phys. Rev.}
  {\bfseries D85} (2012) 125022}
  [\href{https://arxiv.org/abs/1202.0873}{{\ttfamily 1202.0873}}].

\bibitem{PeskinSchroeder}
M.~E. {Peskin} and D.~V. {Schroeder}, \emph{{An Introduction to Quantum Field
  Theory}}. Westview Press, 1995.

\bibitem{HulthenA}
L.~Hulth{\'e}n, \emph{{{\"U}ber die eigenlosunger der Schr{\"o}dinger-gleichung
  des deuterons}}, {\emph{Ark. Mat. Astron. Fys.} {\bfseries 28 A} (1942) 1}.

\bibitem{HulthenB}
L.~Hulth{\'e}n, \emph{{On the Virtual State of the Deuteron}}, {\emph{Ark. Mat.
  Astron. Fys.} {\bfseries 29 B} (1943) 1}.

\bibitem{AkhiezerMerenkov_sigmaHydrogen}
A.~I. {Akhiezer} and N.~P. {Merenkov}, \emph{{The theory of lepton bound-state
  production}}, \href{https://doi.org/10.1088/0953-4075}{\emph{Journal of
  Physics B Atomic Molecular Physics} {\bfseries 29} (1996) 2135}.

\end{thebibliography}\endgroup
